\documentclass[journal]{IEEEtran}
\usepackage{amsfonts}
\usepackage[para]{threeparttable}
\usepackage{booktabs}
\usepackage{hyphenat}
\usepackage{amssymb}
\usepackage{amsmath}
\usepackage{stfloats}

\usepackage{graphicx}
\usepackage{placeins}
\usepackage{float}
\usepackage{subfigure}
\usepackage{array}
\usepackage{algorithm}
\usepackage{algorithmic}
\usepackage{bm}
\usepackage[square, comma, sort&compress, numbers]{natbib}
\usepackage{hyperref}

\usepackage{color}
\usepackage{colortbl}
\usepackage{multirow}

\ifCLASSINFOpdf
\else
\fi

\begin{document}
\title{Line-of-Sight MIMO for High Capacity Millimeter Wave Backhaul in
	FDD Systems}

\author{Ye~Xue$^\dagger$,
        Xuanyu~Zheng$^\dagger$,
        and~Vincent~Lau% <-this % stops a space
\thanks{$^\dagger$  These two authors contribute equally to the work.}% <-this % stops a space
\thanks{Y. Xue, X. Zheng and V. Lau  are with the Hong Kong University of Science and Technology, Hong Kong, China}}% <-this % stops a space

\maketitle

\begin{abstract}
Wireless backhaul is considered to be the key part of the future wireless
network with dense small cell traffic and high capacity demand. In
this paper, we focus on the design of a high spectral efficiency line-of-sight
(LoS) multiple-input multiple-output (MIMO) system for millimeter
wave (mmWave) backhaul using dual-polarized frequency division duplex
(FDD). High spectral efficiency is very challenging to achieve for
the system due to various physical impairments such as phase noise
(PHN), timing offset (TO) as well as the poor condition number of
the LoS MIMO. In this paper, we propose a holistic solution containing
TO compensation, PHN estimation, precoder/decorrelator optimization
of the LoS MIMO for wireless backhaul, and the interleaving of each
part. We show that the proposed solution has robust performance with
end-to-end spectral efficiency of 60 bits/s/Hz for 8x8 MIMO.
\end{abstract}

\begin{IEEEkeywords}
Line-of-Sight MIMO, Wireless backhaul, FDD, timing synchronization,
	phase noise
\end{IEEEkeywords}

% \fi
\IEEEpeerreviewmaketitle

\section{Introduction}
Dense small cells and femtocells have been proposed to enhance the
spatial reuse and boost the capacity of hot spots in future wireless
networks \cite{bhushan2014network,wang2014cellular}. As a result,
future wireless networks may comprise a substantial number of small
base stations, and the backhaul transmission will be a critical capacity
and cost bottleneck. In this paper, we focus on designing a very high
spectral efficiency mmWave backhaul transmission using  MIMO technology
for future wireless networks.

MIMO has been widely used in 3GPP long term evolution (LTE) and 5G
systems as a key technique to enable the capacity requirement of the
wireless access network. Leveraging the rich scattering in the non-line-of-sight
(NLoS) fading channels, spatial multiplexing for MIMO systems has
been widely studied in \cite{zheng2003diversity}. However, there
are various technical challenges in adopting the MIMO technique for
improving the spectral efficiency of the wireless backhaul. First,
in the context of wireless backhaul, the propagation channel is primarily
LoS due to the high carrier frequency together with the narrow beamwidths
being used. With limited scattering in the propagation environment,
the LoS MIMO channel responses can be highly correlated, leading to
a rank deficient channel matrix. Nevertheless, by optimizing the antenna
placements, the capacity and rank of the LoS MIMO channel can be improved
with a large separation between the antennas in the arrays \cite{8690668,4155681}.
However, it is costly to install arrays with large antenna-separation,
especially for MIMO systems. This poses the technical challenge in
adopting the MIMO technique for improving the spectral efficiency
of the wireless backhaul since low antenna-separation arrays will
lead to a poor condition number of the LoS MIMO channel and the spatial
multiplexing benefit can be jeopardized.

The second challenge is brought by the existence of physical impairments
in the system, such as TO and PHN. The presence of timing offsets
among the transmit and receive antenna will bring severe inter-symbol
interference (ISI), which will degrade the performance of the wireless
backhaul. As a result, this poses a very stringent requirement for
accurate timing synchronization. Additionally, PHN in MIMO systems
will bring two penalties in the MIMO system, namely the $\emph{demodulation penalty}$
caused by phase distortion to signal constellation and the $\emph{multi-access interference (MAI) penalty}$
induced by the coherency loss of the precoder and decorrelator.\footnote{The precoder and decorrelator are used to mitigate interference between
	spatially multiplexed data streams. The precoder and decorrelator
	were designed based on the estimated CSI at the beginning of a frame
	and were fixed throughout the frame. However, due to PHN, there is
	an increasing mismatch between the precoder/decorrelator and the effective
	channel (incorporating the PHN) and hence, the MAI increases.} These physical impairments usually exist in practical wireless backhaul,
the effect of which is more severe in mmWave systems \cite{song2018design}
due to the high carrier frequency. Unlike wireless access applications,
the target spectral efficiency of wireless backhaul is very high.
These physical impairments can be the performance bottleneck for wireless
backhaul and pose a very stringent requirement for accurate estimation
and compensation of the TO and PHN. Unfortunately, the spatial multiplexing
in MIMO systems will cause severe MAI, which is a huge hurdle for
the compensation of these impairments. Furthermore, dual-polarization
is usually adopted in wireless MIMO backhaul \cite{camarchia2016electronics}
to overcome the space limitation, which allows two orthogonal streams
from each dipole antenna to travel in the same bandwidth at the same
time (one vertically (V mode) and one horizontally (H mode)). In such
a dual-polarized system, high cross-polarization discrimination (XPD)
of each dipole antenna is required to reduce the interference between
cross-polar link transmission. However, high XPD together with severe
MAI will lead to inaccurate cross-polar link measurements for the
TO and PHN, making it more challenging to estimate and compensate
these physical impairments.\footnote{The cross-polar link represents the data link between the V mode (H
	mode) at the transmitter and the H mode (V mode) at the receiver in
	a dual-polarized MIMO (See Fig.\ref{fig:Illustration-of-antenna}),
	which is very weak in high XPD systems and will be corrupted by the
	MAI caused by the signal from other links.} Thus, a novel and holistic solution considering the physical impairments
for mmWave LoS MIMO backhaul with dual-polarization is required. 

There are many existing works that considered different subsets of
the aforementioned physical impairments \cite{peterson1995introduction,mahesh2012fractional,nasir2013optimal,kung2014robust,mengali2013synchronization,zhao2006novel,godtmann2007iterative,hadaschik2005improving,mehrpouyan2012joint,huang2015phase,lee2017channel,song2018design}.
TO estimation has been well studied in \cite{mahesh2012fractional,nasir2013optimal,kung2014robust}.
In \cite{mahesh2012fractional}, the authors propose a maximum likelihood
(ML) estimator to the TO in the MIMO system with an impractical assumption
that the TOs are identical across all the antennas. \cite{nasir2013optimal,kung2014robust}
propose estimation methods of different TOs over antennas. However,
there are strict assumptions of TO in these works. Specifically, \cite{nasir2013optimal}
assumes TOs are small and within one symbol time and \cite{kung2014robust}
only considers TOs that are multiples of symbol time. In addition,
these works only estimate the sum-offsets \footnote{Sum-offsets are the effective TOs in the transmitter-receiver links.
	For example, in an 8x8 MIMO, we have altogether 16 unknown TOs for
	all the antennas, but these works only estimate the 64 combinations
	of the sum-offsets between the transmitter and receiver antennas.
	There are only 16 freedoms in these 64 sum-offsets, but such information
	is not exploited in these works. } and the preamble used in this work may fail due to the severe MAI
induced by the high XPD in the dual-polarized LoS MIMO channel. 

PHN estimation algorithms in single-input single-output (SISO) systems
are widely studied in \cite{mengali2013synchronization,zhao2006novel,godtmann2007iterative,amblard2003phase}.
However, these methods cannot be applied to MIMO systems since the
received signal at each antenna will be influenced by multiple PHNs
in the transmitters, which needs to be jointly estimated. For MIMO
systems, PHN estimation and compensation algorithms are proposed in
\cite{hadaschik2005improving,mehrpouyan2012joint,huang2015phase}
with data-aided and decision-directed sum-PHNs\footnote{Sum-PHNs are the effective PHNs in the transmitter-receiver links.
	For example, in an 8x8 MIMO, we have altogether 16 unknown PHNs for
	all the antennas, but these works only estimate the 64 combinations
	of the sum-PHNs between the transmitter and receiver antennas. There
	are only 16 freedoms in these 64 sum-PHNs, but such information is
	not exploited in these works. } estimators. However, \cite{hadaschik2005improving} assumes that
the channel is perfectly known at the receiver and \cite{mehrpouyan2012joint,huang2015phase}
have ignored the effect of TOs at the transceiver. In addition, the
high XPD in dual-polarized MIMO is not considered, hence, the sum-PHNs
extracted from the estimated channel will have large estimation errors
for transceiver links with a small amplitude (e.g., the cross-polar
links). In these works, after estimating the sum-PHNs, the authors
design the decorrelator to equalize the aggregated channel and sum-PHN
matrix. These compensation schemes impose a heavy computational burden
at the receiver since the PHN is varying over time and the decorrelator
needs to be updated along with the variations of the PHN. In \cite{song2018design},
the authors propose a method to estimate and compensate the per-antenna
PHN for MIMO, but they do not track the time-varying PHN and they
ignore the interplay with the precoder and decorrelator. Hence, existing
works are not applicable when various impairments are considered in
mmWave LoS MIMO. 
The precoder and decorrelator are key components in MIMO systems to
achieve high spectral efficiency for wireless backhaul, since they
enable stable multi-stream transmissions. There are a lot of existing
works on MIMO precoder/decorrelator design \cite{palomar2003unified,sampath2001generalized,shi2011iteratively}.
\cite{palomar2003unified} proposes a unified linear transceiver design
framework, in which the optimal decorrelator is fixed as the Wiener
filter and the design problem can be formulated as convex optimization
problems in terms of precoder under different design criteria. In
\cite{sampath2001generalized,shi2011iteratively}, the authors consider
the weighted MMSE optimization and propose low complexity algorithms
based on alternative optimization. However, in all these works, the
physical impairments such as TO and PHN have been ignored. This may
be justified for wireless access applications but these physical impairments
cannot be ignored for wireless backhaul applications due to the very
high target spectral efficiency. Recently, many works have considered
the physical impairments issue. Hardware impairments aware (HIA) MIMO
transceivers are proposed in \cite{taghizadeh2018hardware2,xia2015hardware,taghizadeh2018hardware}.
However, the effect of physical impairments is simply modeled as additive
Gaussian noise to each antenna, which is an over-simplification of
the impairments due to the TO and the PHN in the MIMO systems. In
summary, these existing solutions cannot be applied in our case because
of very different target spectral efficiency and the practical considerations
of the LoS MIMO.

In this paper, we adopt a holistic approach and propose a practical
solution for a dual-polarized LoS MIMO in mmWave backhaul addressing
the aforementioned physical impairments. The solution is not a trivial
combination of existing techniques as each component is inter-related.
Based on the proposed solution, we can achieve very high spectral
efficiency (e.g., 60 bits/s/Hz with 8x8 MIMO) and fully unleash the
potential of the LoS MIMO in mmWave backhaul applications. The following
summarizes our contributions.

\begin{itemize}
	\item \textbf{Decentralized Spatial Timing Estimation and Compensation:
	}We propose a low complexity spatial timing estimator that has a similar
	complexity to the cross-correlation approach \cite{peterson1995introduction,kung2014robust}
	but it is capable of utilizing spatial information across different
	antennas to estimate the per-antenna TO. The proposed spatial TO estimator
	only requires local information and hence it can be implemented separately
	at the transmitter and receiver. To overcome the strong MAI induced
	by spatial multiplexing of dual-polarized LoS MIMO channels, we propose
	new preamble sequences with improved auto-correlation and cross-correlation
	rejection. Based on this, we propose a decentralized timing compensation
	scheme where the transmitter and receiver compensate for the TO based
	only on local information without any explicit signaling.
	\item \textbf{Decentralized Phase Noise Estimation and Compensation with
		Decision Feedback. }We propose a low complexity per-antenna PHN estimation
	and compensation scheme, which enables compensation for both the phase
	distortion to the received symbols and the MAI caused by the loss
	of coherence of the precoder and decorrelator due to the drifting
	of PHN.\footnote{Note that such per-antenna compensation is not possible if one uses
		conventional PHN estimators, which only estimate and compensate the
		effective sum-PHN.} To reduce the pilot overhead, we adopt the decision feedback and
	regression-based fusion to enhance the PHN estimation quality. The
	proposed PHN estimator has low complexity and requires local information
	only. Based on this, we propose a decentralized PHN compensation scheme,
	which compensates the per-antenna PHN locally at the transmitter and
	the receiver. 
	\item \textbf{Robust MIMO Precoder and Decorrelator Design}: We exploit
	the MIMO precoder and decorrelator to suppress the inter-symbol interference
	(ISI) and MAI induced by the physical impairments of the TO and the
	PHN. We show that the precoder and decorrelator can significantly
	alleviate the requirement of the TO compensation and PHN compensation
	to achieve very high spectral efficiency for mmWave backhaul applications.
	The design is formulated as a nonconvex optimization problem. By exploiting
	structures in the ISI and MAI, we transform the problem into a tractable
	form and propose a low complexity solution using alternative optimization
	techniques.
\end{itemize}
This paper is organized as follows. In Section \ref{sec:System-Model},
we present the system model, including the mmWave dual-polarized LoS
MIMO channel model, the TO model, and the PHN model, as well as the
data path in the transmitter and receiver. In Section \ref{sec:Joint-Timing-Estimation},
\ref{sec:Phase-Offset-Estimation} and \ref{sec:Robust-MIMO-Precoder},
the proposed timing synchronization, PHN estimation and compensation,
as well as the procoder and decorrelator scheme, respectively, are
illustrated. The numerical simulation results and the corresponding
discussions are provided in Section \ref{sec:Simulations-and-Discussions}.
Finally, Section \ref{sec:Conclusion} summarizes the whole work.

\textbf{Notations:} In this paper, lowercase and upper bold face letters
stand for column vectors and matrices, respectively. The operations
$\left(\cdot\right){}^{T}$ and $\left(\cdot\right)^{H}$ are respectively,
the operations of transpose and conjugate transpose. The entry in
the $i$-th row and $j$-th column of matrix $\mathbf{A}$ is $\left[\mathbf{A}\right]{}_{i,j}$
while the $n$-th element of vector $\mathbf{a}$ is $a_{n}$. $a^{*}$
represents the complex conjugate of $a$. $\mathbf{I}_{N}$ is the
$N\times N$ identity matrix and $\mathbf{1}_{N}$ denotes the all
one vector of dimension $N$. The norms $||\cdot||_{2}$, $|\cdot|$
and $||\cdot||_{F}$ are respectively, the $L_{2}$, $L_{1}$ and
Frobenius norm. $\odot$ and $\otimes$ represents the Hadamard product
and Kronecker product, respectively. $e^{^{j\left(\cdot\right)}}$
is an elementwise operator for vector or matrix input. $\mathcal{CN}(\mu,\sigma^{2})$
denotes the Complex Gaussian distribution with mean, $\mu$ and variance
$\sigma^{2}$. $\mathrm{diag}\left(\mathbf{a}\right)$ denotes a diagonal
matrix whose diagonal elements are filled with elements of vector
$\mathbf{a}$, and $\mathrm{vec}\left(\cdot\right)$ is the vectorization
operator. Finally, $g(\cdot)*f(\cdot)$ denotes the convolution of
$g(\cdot)$ and $f(\cdot)$. 

\section{System Model \label{sec:System-Model}}

In this paper, we consider mmWave MIMO wireless backhaul in a dual-polarized
FDD system. Both the transmitter and the receiver comprise an antenna
array mounted on a pole with a primarily LoS channel in between. Each
of the transmit (and receive) antennas has a local oscillator that
is loosely synchronized to a master control unit. The illustration
of the system is shown in Fig.\ref{fig:Illustration-of-antenna}.
We shall elaborate on each part of this system below. 

\subsection{LoS MIMO Channel Model }

In mmWave backhaul, a terrestrial link usually exists. Therefore,
we consider the Rummler model \cite{rummler1979new}  in this work,
which is an LoS propagation model with a single  NLoS path caused by the terrestrial reflection between two fixed antenna towers.
For a single-input-single-output (SISO) system, the impulse response
of the Rummler model can be specified as

\begin{equation}
h\left(t\right)=\delta\left(t\right)+\beta e^{j2\pi f_{0}\tau^{d}}\delta\left(t-\tau^{d}\right),\label{eq:mutimod}
\end{equation}
where the first term represents the LoS path and the second term represents
the NLoS path, $\tau^{d}$ is the propagation delay associated with
the difference in the propagation time between the LoS path and the
NLoS path and is assumed to be within one symbol time. $f_{0}$ denotes
the notch frequency, which will be anywhere in the spectral efficiency.
We consider the minimum-phase case of the Rummler model ($\beta\leq1$),
where we set the channel gains $\beta=1-10^{-\frac{\rho}{20}}$, with
$\rho$ denoting the notch depth, which relates the power of the LoS
path to that of the NLoS paths.

\begin{figure}
	
	\centering{}
	\includegraphics[width=0.5\textwidth]{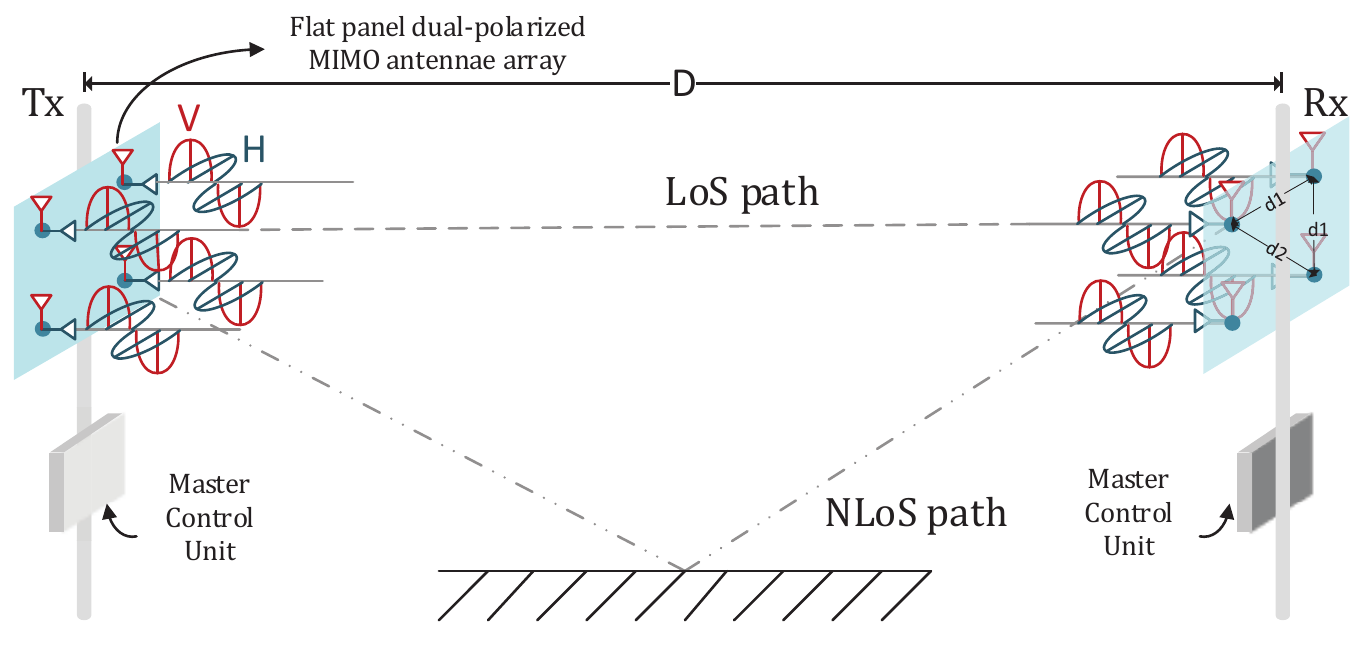}\caption{\label{fig:Illustration-of-antenna}Illustration of antenna arrays
		and geometry of an $8\times8$ LoS dual-polarized MIMO system. Flat-panel
		dual-polarized MIMO antenna arrays are applied with antenna spacing
		$d1$, $d2$, where $d2=\sqrt{2}d1$. The distance between the transmitter
		and the receiver is $D$.}
\end{figure}

Generalizing the SISO Rummler model to a dual-polarized \footnote{In this paper, we consider the H mode and V mode of one dipole antenna
	as two different antennas. For example, in Fig.\ref{fig:Illustration-of-antenna},
	there are 4 dipole antennas at each side of the transmitter and the
	receiver, and the system is considered to be a $8\times8$ MIMO system. } $N\times M$ MIMO system, the channel model is given by
\begin{align}
\tilde{\mathbf{H}}(t)&=\mathbf{H}^{LoS}(t)+\mathbf{H}^{NLoS}(t),\nonumber\\
&=\mathbf{H}^{LoS}(t)+\mathbf{\boldsymbol{\beta}}\odot\mathbf{R}\odot\mathbf{H}^{LoS}(t-\tau^{d})
\label{eq:MIMomodel}
\end{align}
where $\boldsymbol{\beta}$ and $\mathbf{R}$ is the random magnitude attenuation
and the random phase rotation matrix, respectively, caused by reflection.Since  $\boldsymbol{\beta}$ and $\mathbf{R}$  are multiplied element-wisely to the LoS channel response, the expression for the NLoS propagation in (\ref{eq:MIMomodel}) can represent any NLoS propagation.
The LoS channel $\mathbf{H}^{LoS}(t)$ is highly deterministic \cite{song2013millimeter}
and can be modeled as

\begin{equation}
\mathbf{H}^{LoS}=\mathbf{H}_{xp}\otimes\mathbf{J}_{\ensuremath{\frac{N}{2}}\ensuremath{\times}\ensuremath{\frac{M}{2}}}\odot\mathbf{H}_{A},
\end{equation}
where $\mathbf{H}_{xp}\in\mathbb{C}^{2\times2}$ contains the cross-polar
gains, $\mathbf{\mathbf{J}}_{\ensuremath{\frac{N}{2}}\ensuremath{\times}\ensuremath{\frac{M}{2}}}$
is the $\ensuremath{\frac{N}{2}}\ensuremath{\times}\ensuremath{\frac{M}{2}}$
all ones matrix and $\mathbf{H}_{A}\in\mathbb{C}^{\text{\ensuremath{N}\ensuremath{\times}\ensuremath{M}}}$
is the array response matrix between transmitter and receiver with
$N$ and $M$ flat-panel dual-polarized MIMO antenna arrays. Note
that in this paper, we assume spherical curvature of the propagating
waves, thus the array response $\mathbf{H}_{A}$ is given by

\begin{align*}
[\mathbf{H}_{A}]_{i,j} & =e^{-j\frac{2\pi}{\lambda}d_{i,j}},
\end{align*}
where $d_{i,j}$ is the distance between the $i$-th receiver antenna
and the $j$-th transmit antenna. A similar propagation assumption
and the array response matrix can be found in \cite{song2018design},
in which the dual-polarized array is not considered. 

\begin{figure}
	\begin{centering}
		\includegraphics[width=0.4\textwidth]{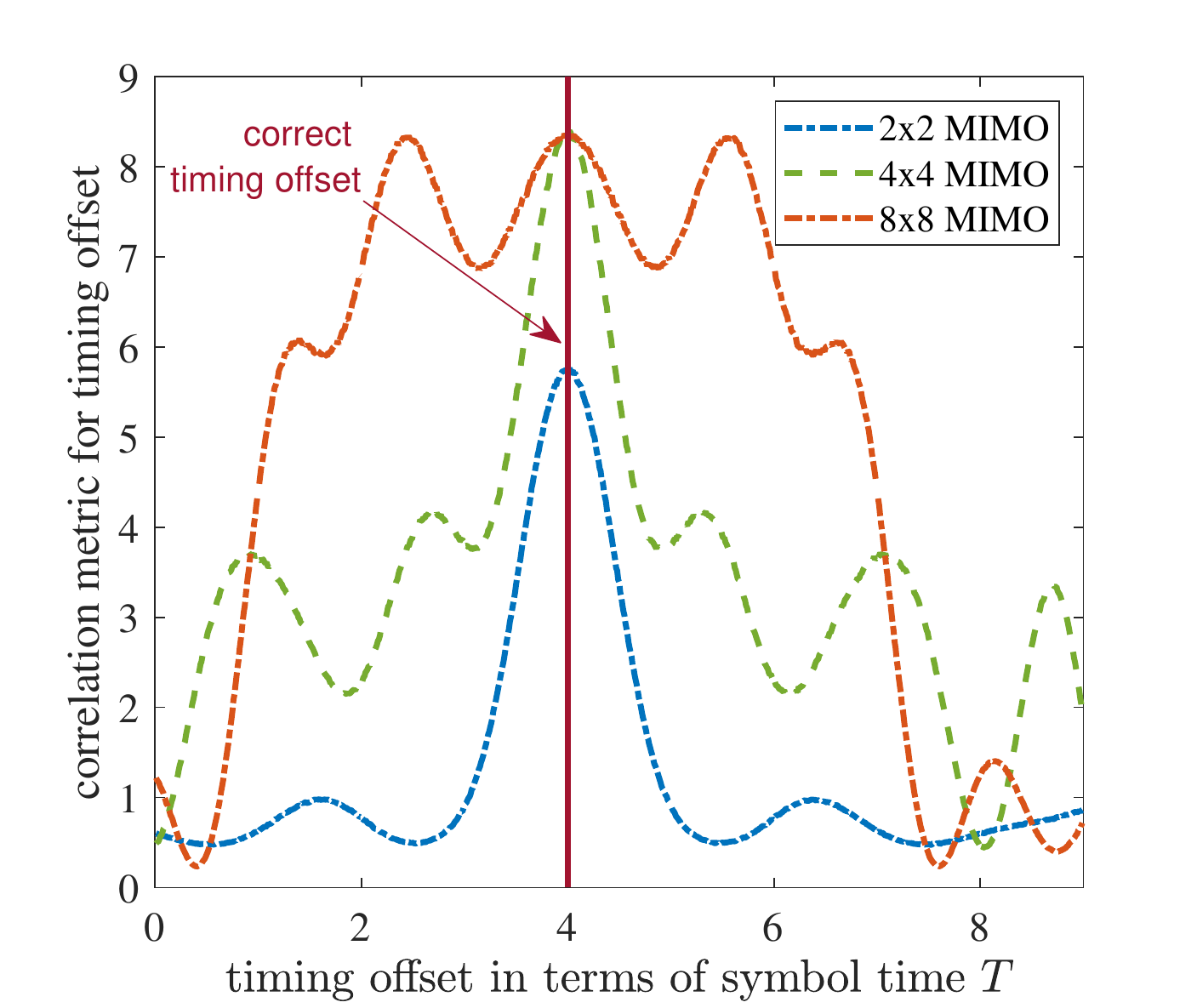}
		\par\end{centering}
	\caption{Correlation metric for TO estimation in a cross-polar link for different
		numbers of antennas. The traditional ZC sequence
		is used as the preamble sequence for TO estimation.
		$\mathrm{XPD}$ is set to a typical value of $20$
		dB, and the TO is set to be 4 times the symbol time.
		\label{fig:Correlation-ZC}}
\end{figure}

With dual-polarized MIMO antennas, XPD of the antennas has a significant
influence on the system performance. In general, high antenna XPD
is desired to leverage the benefits of dual-polarization \cite{ibnkahla2008adaptive,perez2004performance}.
However, high XPD also causes very weak cross-polar transmission links,
i.e., the link between the horizontal mode to the vertical mode of
each dipole antenna. This will cause a great challenge for the estimation
of the TO and PHN since the physical impairments are different on
each antenna and the estimation for these physical impairments requires
measurements from all the transceiver links. Moreover, the intensity
of the MAI will increase as the number of antennas of the MIMO system
increases, which will further distort the cross-polar measurements.
For example, in an $8\times8$ MIMO with a typical $\mathrm{XPD}=20$
dB, the power of the MAI will be $26$ dB of the desired signal for
a cross-polar link. We show this influence on TO estimation in Fig.
\ref{fig:Correlation-ZC} with the timing correlation metric in a
cross-polar link using the traditional Zadoff--Chu (ZC) sequence.
It is observed that as the number of antennas increases, the severe
MAI will induce several peaks of similar intensity in the timing correlation
metric. As a result, there might be large timing estimation errors
due to the \emph{false peaks} in the metric. 

\subsection{Timing offset Model}

In mmWave MIMO wireless backhaul, the transmit antennas are not collocated
and hence the clock of each transmit antenna/receive antenna is only
roughly synchronized to a master clock. Let $\tau_{j}^{tx}=\frac{1}{T}|\text{offset}_{j}^{tx}-\text{offset}_{1}^{tx}|\text{ }$and
$\tau_{i}^{rx}=\frac{1}{T}|\text{offset}_{i}^{rx}-\text{offset}_{1}^{tx}|$
denote the normalized TO of the clock at the $j$-th transmit antenna
and $i$-th receive antenna, respectively, w.r.t. the symbol duration
$T$. Without loss of generality, we assume $\tau_{1}^{rx}=0$ and
TOs $\tau_{j}^{tx}$ and $\tau_{i}^{rx}$ are independent quasi-static
random processes\footnote{ In practice, the drift in the oscillators is a slowly varying process
	with coherence time in the order of minutes or hours \cite{sadler2006synchronization}.} uniformly distributed in $(0,\tau_{max})$, where $\tau_{max}>1$
is the maximum timing delay that can occur in the system. We denote
$\boldsymbol{\tau}^{tx}=\left[\tau_{1}^{tx},...,\tau_{M}^{tx}\right]^{T}$
and $\boldsymbol{\tau}^{rx}=\left[\tau_{1}^{rx},...,\tau_{N}^{rx}\right]^{T}$
as the TO vector for the transmitter and receiver, respectively, and
let $\boldsymbol{\tau}=\left[\left(\boldsymbol{\tau}^{rx}\right)^{T},\left(\boldsymbol{\tau}^{tx}\right)^{T}\right]^{T}$.

\subsection{Phase Noise Model}

In this paper, we consider the independent phase noise $\theta_{j}^{tx}\left(n\right)$
on each transmit antenna and $\theta_{i}^{rx}\left(n\right)$ on each
receive antenna. For free-running oscillators, the discrete time PHNs
$\theta_{j}^{tx}\left(n\right)$ and $\theta_{i}^{rx}\left(n\right)$
for $j=1,...,M$ and $i=1,...,N$ can be modeled as a Wiener process
\cite{chorti2006spectral,hadaschik2005improving,mehrpouyan2012joint},
which are given by
\begin{align}
\theta_{j}^{tx}\left(n\right) & =\theta_{j}^{tx}\left(n-1\right)+\Delta_{j}^{tx}\left(n\right),\nonumber \\
\theta_{i}^{rx}\left(n\right) & =\theta_{i}^{rx}\left(n-1\right)+\Delta_{i}^{rx}\left(n\right).\label{eq:PHN_Model}
\end{align}
The terms $\Delta_{j}^{tx}\left(n\right)$ and $\Delta_{i}^{rx}\left(n\right)$
are random phase innovations for the oscillators at each sample, assumed
to be white real Gaussian processes with $\Delta_{j}^{tx}\left(n\right)\sim\mathcal{N}\left(0,\sigma_{\Delta_{j}^{tx}}^{2}\right)$
and $\Delta_{i}^{rx}\left(n\right)\sim\mathcal{N}\left(0,\sigma_{\Delta_{i}^{rx}}^{2}\right)$,
respectively. $\sigma_{\Delta_{j}^{tx}}^{2}$ and $\sigma_{\Delta_{i}^{rx}}^{2}$
stands for the variance of the innovations at the $j$-th and $i$-th
transmit and receive antennas, respectively, which are given by \cite{mehrpouyan2012joint,schenk2008rf}
\begin{align}
\sigma_{\Delta_{j}^{tx}}^{2} & =2\pi c_{j}^{tx}T_{s},\nonumber \\
\sigma_{\Delta_{i}^{rx}}^{2} & =2\pi c_{i}^{rx}T_{s},\label{eq:PHNvariance}
\end{align}
where $c_{j}^{tx}$ and $c_{i}^{rx}$ denote the one-sided 3 dB bandwidth
of the Lorentzian spectrum of the oscillators at the $j$-th and $i$-th
transmit and receive antennas, respectively, and $T_{s}$ is the sampling
time. We assume that $\sigma_{\Delta_{j}^{tx}}^{2}$ and $\sigma_{\Delta_{i}^{rx}}^{2}$
are known at the receiver since they are dependent on the oscillator
properties. 

\subsection{Transmit and Receive Data Path}
\begin{figure}
	\begin{centering}
		\includegraphics[width=0.5\textwidth]{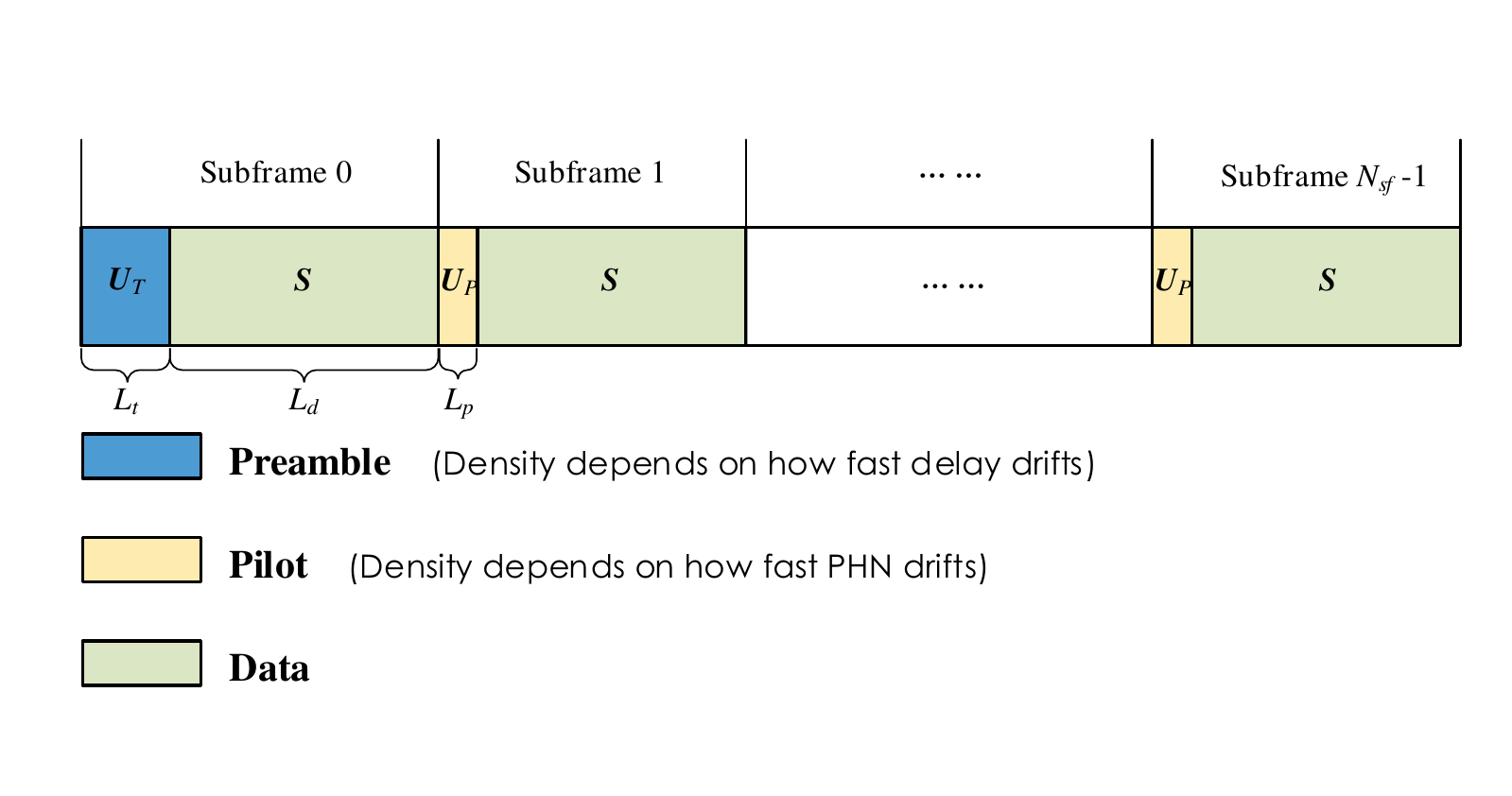}
		\par\end{centering}
	\caption{Frame structure for the transmission of preamble, pilot, and data
		symbols from one transmitter.\label{fig:Timing-diagram-frame}}
\end{figure}

We consider a frame structure, illustrated in Fig. \ref{fig:Timing-diagram-frame},
in this paper. There is a preamble sequence $\mathbf{U}_{T}=\left[\mathbf{a}_{1},...,\mathbf{a}_{M}\right]^{T}\in\mathbb{C}^{M\times L_{t}}$
with $L_{t}$ symbols at the beginning of each frame for timing synchronization
and channel estimation. Specifically, since timing offset is a slowly
varying process, we use the preamble of the first frame (initial frame)
to perform timing synchronization, while preambles of the rest frames
are used for channel estimation. Each frame consists of $N_{sf}$
subframes, and within each subframe, there is a data section with
$L_{d}$ symbols, denoted by $\mathbf{S}=\left[\mathbf{s}_{1},...,\mathbf{s_{M}}\right]^{T}\in\mathbb{C}^{M\times L_{d}}$,
where $\mathbf{s}_{j}\in\mathbb{C}^{L_{d}\times1}$ is the transmitted
data symbols at $j$-th transmit antenna. In addition, there is a
pilot sequence $\mathbf{U}_{P}=\left[\mathbf{b}_{1},...,\mathbf{b}_{M}\right]^{T}\in\mathbb{C}^{M\times L_{p}}$
with length $L_{p}$ at the beginning of each subframe for the estimation
and tracking of the PHN over the entire frame. To summarize, the first
subframe of each frame has the length of $L_{t}+L_{d}$ symbols, while
the other subframes are of length $L_{p}+L_{d}$ symbols. As a result,
the overall pilot and preamble overhead is given by $\frac{L_{t}+\left(N_{sf}-1\right)L_{p}}{N_{sf}L_{d}}$.
In practice, the period of the preamble is comparable to the coherence
time of the channel (which is $10$\textasciitilde$100${} ms for
fixed-point wireless backhaul applications \cite{hur2013millimeter}).
The period of the pilot is more frequent as it has to track the variation
of the PHN $\theta_{j}^{tx}\left(n\right)$ and $\theta_{i}^{rx}\left(n\right)$.
For a typical oscillator at $f_{c}=28$ GHz, the effective bandwidth
of the power spectral density (PSD) of the PHN is $10^{4}$ Hz \cite{kuo2017phase}
and hence, the coherence time of the PHN is around $0.1$ ms. Consequently,
a pilot density of $100$ pilots per frame would be sufficient. Hence,
for a system with a symbol rate of $25$ M/s, the pilot and preamble
overhead is around 5\% for $L_{t}=256$, $L_{p}=64$, $L_{d}=1280$
and $N_{sf}=100$, which is quite low. 

\begin{figure*}
	\centering
	\includegraphics[scale=0.43]{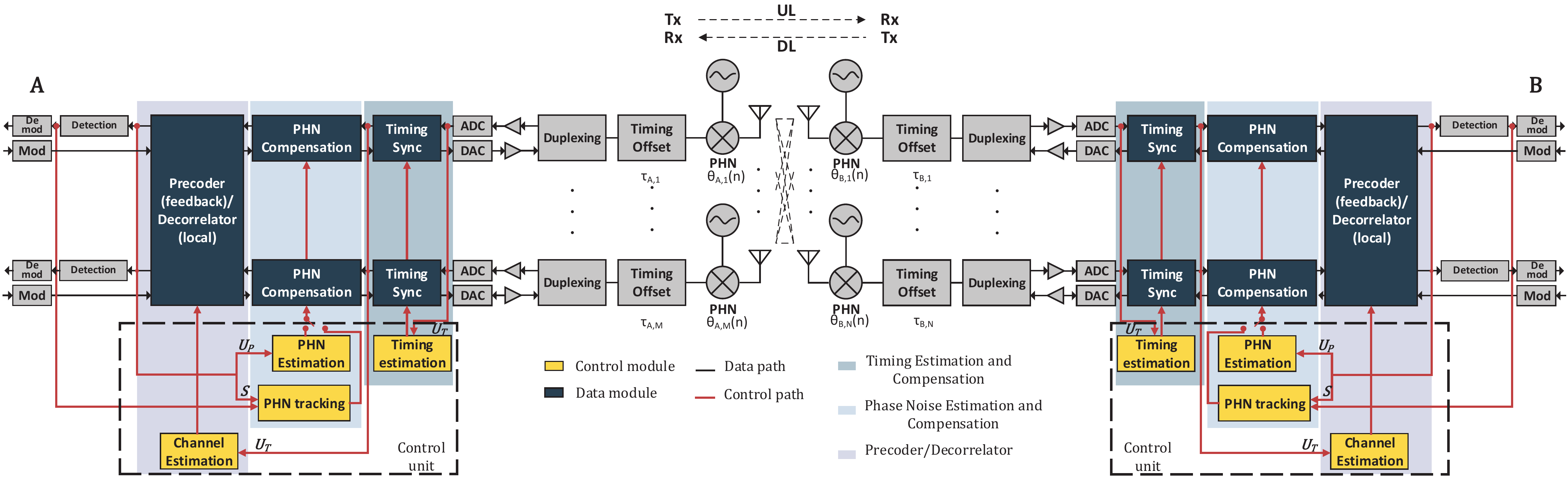}
	\caption{Datapath for the proposed mmWave MIMO backhaul in FDD system.}
	\label{fig:Data_path}
	
\end{figure*}

Figure.\ref{fig:Data_path} illustrates the uplink and downlink datapath
of the transmitter and receiver for the wireless backhaul between
two base stations (BSs) A and B. We consider an FDD system, and the
transmitter and the receiver of a site share a common oscillator.
Hence, the PHN and the TO during downlink for BS A (BS B) as a receiver
(transmitter) are identical to the PHN and the TO during uplink for
BS A (BS B) as a transmitter (receiver), i.e., in Fig. \ref{fig:Data_path}
we have $\theta_{A,j}^{UL}\left(n\right)=\theta_{A,j}^{DL}\left(n\right)$,
$\tau_{A,j}^{UL}=\tau_{A,j}^{DL}$, $\forall j$ and $\theta_{B,i}^{UL}\left(n\right)=\theta_{B,i}^{DL}\left(n\right)$,
$\tau_{B,i}^{UL}=\tau_{B,i}^{DL}$, $\forall i$. To avoid the abuse
of notation, we assume $\theta_{j}^{tx}(n)=\theta_{A,j}^{UL}\left(n\right)$,
$\tau_{j}^{tx}=\tau_{A,j}^{UL}$, $\theta_{i}^{rx}(n)=\theta_{B,i}^{UL}\left(n\right)$,
$\tau_{i}^{rx}=\tau_{B,i}^{UL}$ in the following illustrations.

Let $g\left(t\right)$ be the response of the pulse-shaping filters
evaluated at $t$. The equivalent channel for the $j$-th transmit
antenna and the $i$-th receive antenna is given by
\begin{equation}
h_{i,j}^{\mathbf{\boldsymbol{\tau}}}\left(t\right)=\left[\tilde{\mathbf{H}}\left(t\right)\right]_{i,j}*g\left(t-\left(\tau_{j}^{tx}+\tau_{i}^{rx}\right)T\right),\label{eq:equivalent_ch}
\end{equation}
where the superscript $\boldsymbol{\tau}$ means the variable is parameterized
by $\boldsymbol{\tau}$. Let $\mathbf{u}_{j}=\left[u_{j}(0),u_{j}(1),...,u_{j}(L-1)\right]^{T}$
denote the complex-valued symbol sequence of length $L$ transmitted
by the $j$-th transmitter, which can be a preamble sequence, pilot
sequence or data. Suppose the received waveform is sampled at a rate
of $Q$ samples per symbol (i.e., $T_{s}=\frac{T}{Q}$), the received
signal of $i$-th receive antenna at the $n$-th sample is given by
\begin{align}
& y_{i}\left(n\right)\label{eq:Signal-Model}\\
= & \sum_{j=1}^{M}\sum_{k=0}^{L-1}e^{j\left[\theta_{j}^{tx}\left(n\right)+\theta_{i}^{rx}\left(n\right)\right]}h_{i,j}^{\mathbf{\tau}}\left(nT_{s}-kT\right)u_{j}\left(k\right)+v_{i}\left(n\right),\nonumber 
\end{align}

where $k$ is the symbol index, and $v_{i}\left(n\right)$ is the
complex additive white Gaussian noise (AWGN) with zero mean and variance
$\sigma^{2}$, i.e., $v_{i}\left(n\right)\sim\mathcal{CN}(0,\sigma^{2}\text{).}$
Specifically, to show a symbol level signal model, the received signal
at the $n=kQ$-th sample for symbol index $k=0,...,L-1$ is given
by:
\begin{align}
y_{i}\left(kQ\right) & =\underbrace{e^{j\left[\theta_{i}^{tx}\left(kQ\right)+\theta_{i}^{rx}\left(kQ\right)\right]}}_{\mathrm{PHN\ distortion}}\left\{ \underbrace{h_{i,i}^{\mathbf{\tau}}\left(0\right)u_{i}\left(k\right)}_{\mathrm{desired\ signal}}\right.\nonumber \\
& +\underbrace{\sum_{k'=1;k'\neq k}^{L-1}h_{i,i}^{\mathbf{\tau}}\left(\left(k-k'\right)T\right)u_{i}\left(k'\right)}_{\mathrm{ISI}}\nonumber \\
& \left.+\underbrace{\sum_{j=1;j\neq i}^{M}\sum_{k'=0}^{L-1}h_{i,j}^{\mathbf{\tau}}\left(\left(k-k'\right)T\right)u_{j}\left(k'\right)}_{MAI}\right\} +\underbrace{v_{i}\left(kQ\right)}_{AWGN}.\label{eq:Symbol-Sig-Model}
\end{align}
From (\ref{eq:Symbol-Sig-Model}), we have the following observations.
First, the PHN will introduce a random phase distortion on the received
signal constellation during the demodulation process. Furthermore,
the MAI and ISI of the spatially multiplexed streams will be mitigated
by the precoder and the decorrelator, which are designed based on
the estimated channel at the beginning of each frame and remain constant
throughout the frame. While the channel will be quasi-static within
a frame, the PHN process will be time-varying and hence the PHN distortion
will induce loss of coherency of the precoder and decorrelator for
symbols within a frame. Second, the impacts of the TO appear in both
the ISI and the MAI terms, as illustrated in (\ref{eq:Symbol-Sig-Model}),
which will increase both types of interference. These physical impairments
may not be a significant performance bottleneck in regular wireless
access applications. However, due to the very high spectral efficiency
target in wireless backhaul applications, they can be a significant
bottleneck. 
\section{Timing Offset Estimation and Compensation\label{sec:Joint-Timing-Estimation}}

We consider the TO issue in this section. Conventional correlation-based
TO estimators correlate the preamble sequences with delayed replicas
of the received samples at each receive antenna, and find the maximum
correlation peak in the timing metric to estimate the sum-offsets
in each transmitter-receiver link. For an $N\times M$ MIMO system,
the correlator can estimate $NM$ sum-offsets for all the transmitter-receiver
links, but there are only $N+M$ degrees of freedom in these sum-offsets
as we only have one unknown TO for each individual antenna to be estimated.
Hence, there are correlations among the $NM$ sum-offsets, and this
correlation can be further exploited to enable a more accurate TO
estimation and compensation for each individual antenna.

Conventionally, the ZC sequence is used as preamble in LTE systems,
serving as the primary synchronization signal (PSS) to extract timing
information. The cross-correlation of the ZC sequence is low for traditional
wireless MIMO applications due to the moderate MAI, thus the accurate
estimations of the sum-offsets can be guaranteed. However, the cross-correlation
of the ZC sequence is not sufficiently low to mitigate the MAI for
a dual-polarized LoS MIMO channel with high XPD, especially for the
cross-polar links, which experience severe MAI. Thus, the traditional
ZC preamble is not enough to isolate MAI in cross-correlation, which
hinders correlator-based timing synchronization from fully utilizing
the spatial information. In addition, the high spectral efficiency
of wireless backhaul poses a very stringent requirement for accurate
timing synchronization, thus preamble sequences with a good auto-correlation
property that is robust to ISI are also required. This motivates us
to design preamble sequences with superior auto/cross-correlation
properties for the TO estimation. 

\subsection{Preamble Sequence Design\label{subsec:Preamble-Sequence-Design}}

\begin{figure}[htbp]
	\centering
	\subfigure[Auto-correlation of $\mathbf{a}_{1}$.\label{fig:Auto-correlation}]{
		\begin{minipage}[t]{0.45\linewidth}
			\centering
			\includegraphics[width=1.5in]{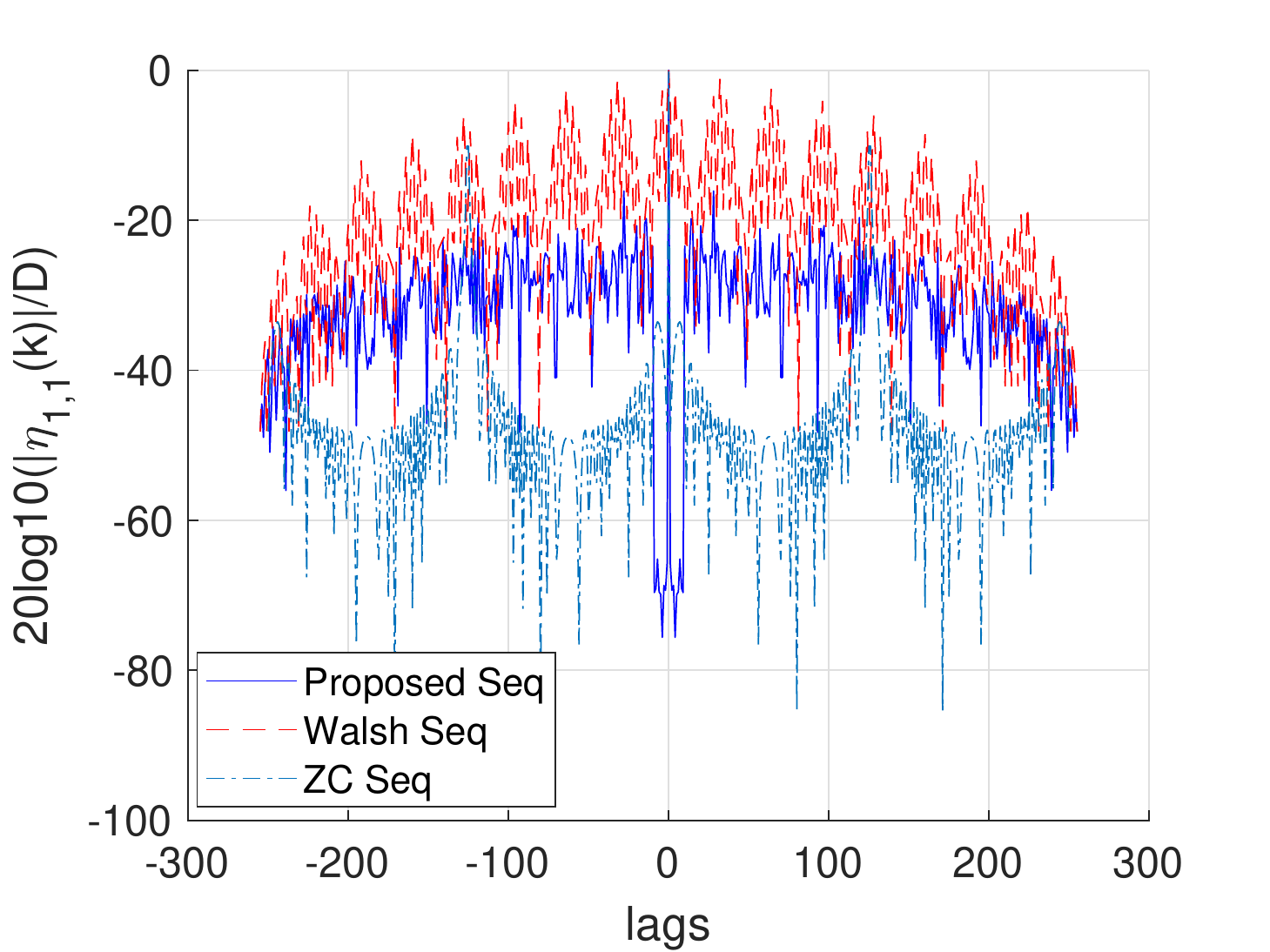}
			%\caption{}
		\end{minipage}%
	}%
	\subfigure[Cross-correlation of $\mathbf{a}_{1}$ and $\mathbf{a}_{2}$.\label{fig:Cross-correlation}]{
		\begin{minipage}[t]{0.45\linewidth}
			\centering
			\includegraphics[width=1.5in]{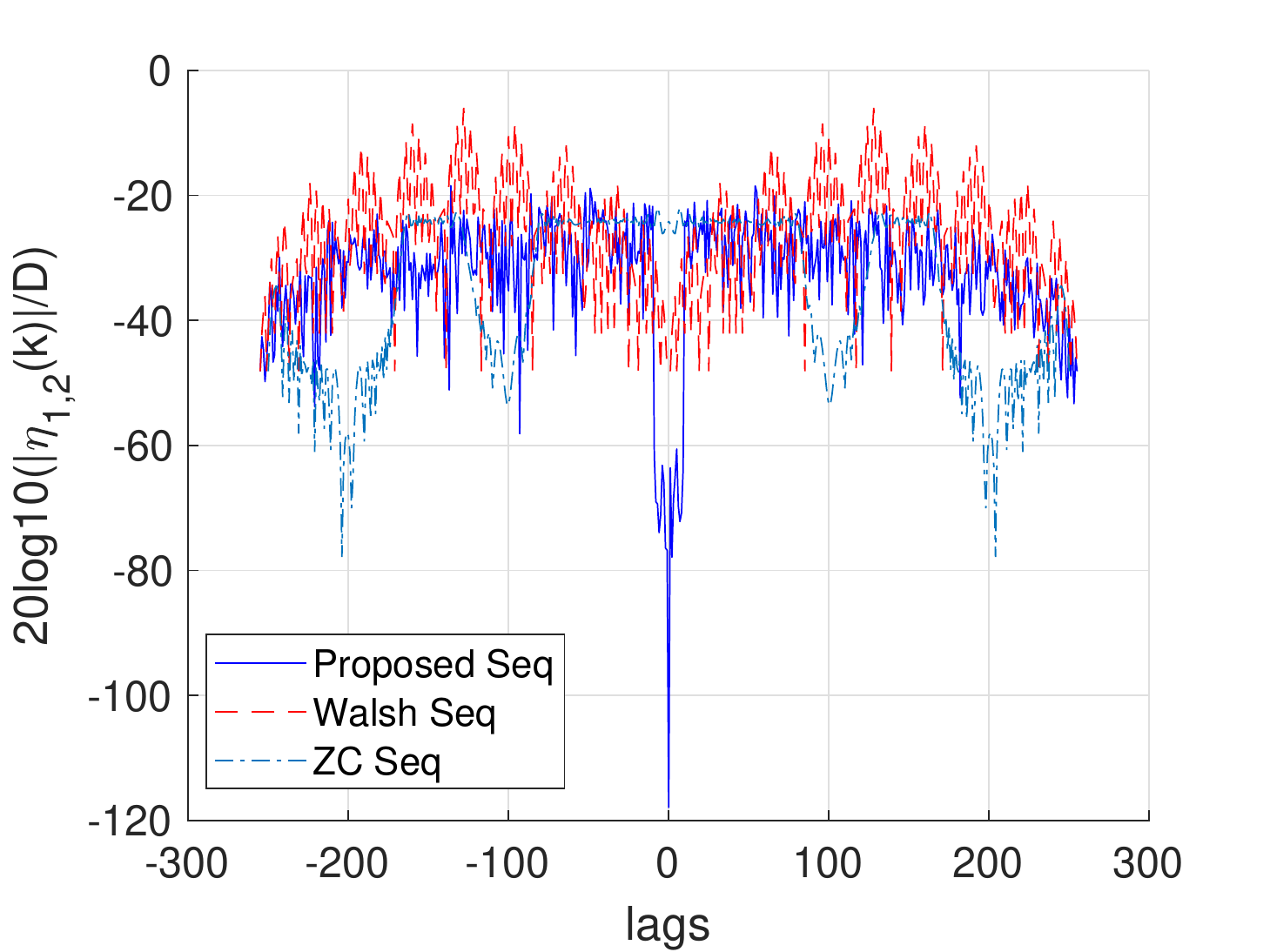}
		\end{minipage}%
	}%
	\centering
	\caption{ Auto/cross correlation of the proposed sequence, the ZC sequence,
		and the Walsh sequence, with $M=8$, $L_{t}=256$ and $\tau_{\mathrm{max}}=5$.
		\label{fig:Auto/cross-correlation}}
\end{figure}

Let $\mathbf{a}_{j}=\left[a_{j}\left(0\right),...,a_{j}\left(L_{t}-1\right)\right]^{T}$
denote the preamble sequence transmitted by the $j$-th transmit antenna.
The cross-correlation of two preamble sequences $\mathbf{a}_{j}$
and $\mathbf{a}_{j'}$ is defined as

\begin{align}
\eta_{j,j'}\left(l\right) & =\sum_{k=0}^{L_{t}-1-l}a_{j}\left(k+l\right)a_{j'}^{*}\left(k\right)=\eta_{j',j}^{*}\left(-l\right),\nonumber \\
& \forall l=0,...,L_{t}-1,\label{eq:cross-correlation-obj}
\end{align}
where $l$ represents the $l$-th lag. Note that (\ref{eq:cross-correlation-obj})
reduces to the auto-correlation of $\mathbf{a}_{j}$ when $j=j'$.
A satisfying set of preamble sequences should have a very low cross-correlation
for all possible lags and a high auto-correlation only when $l=0$,
thus, the correlation peak occurs only when the two sequences are
from the same antenna and are perfectly aligned. However, \cite{he2012waveform}
shows that it is impossible to design such a set. Fortunately, it
is feasible to achieve the required properties in a specific lag interval.
Since the TO will not exceed a maximum value of $\tau_{\mathrm{max}}$
in each transmit and receive antenna, we can seek preamble sequences
with the required properties at the lag interval $0\leq l\leq\lceil2\tau_{\mathrm{max}}\rceil$.
Such a set of sequences can be obtained by solving the following optimization
problems \cite{song2016sequence}:
\begin{align}
\mathrm{min}_{\mathbf{U}_{T}} & \sum_{j=1}^{M}\sum_{j'=1}^{M}\sum_{l=1-L_{t}}^{L_{t}-1}\omega_{l_{f}}\left|\eta_{j,j'}\left(l\right)\right|^{2}-\omega_{0}L_{t}^{2}M\nonumber \\
\mathrm{s.t.} & \left|a_{j}\left(k\right)\right|=1,\ \ k=0,...,L_{t}-1,\ \ j=1,...,M,\label{eq:optSeq}
\end{align}
where $\omega_{l}=\omega_{-l}\geq0$, $l=0,...,L_{t}-1$ are non-negative
weights assigned to different time lags and are defined in our problem
as
\[
\omega_{\pm l}=\left\{ \begin{array}{cc}
1, & 0\leq l\leq\lceil2\tau_{\mathrm{max}}\rceil,\\
0, & \mathrm{otherwise}.
\end{array}\right.
\]
The unimodular constraint in (\ref{eq:optSeq}) ensures the symbols
in preamble sequences have a constant amplitude. The optimization
problem (\ref{eq:optSeq}) is solved via the majorization-minimization
(MM) algorithm and can be implemented efficiently for very long sequences
via fast Fourier transform (FFT) \cite{song2018design}. Fig. \ref{fig:Auto/cross-correlation}
shows the auto/cross-correlation of the proposed preamble sequence,
compared with the ZC and Walsh sequence. It shows that the proposed
sequences provide -70 dB isolation from MAI in cross-correlation and
-70 dB isolation from ISI in auto-correlation at desired lags, which
is sufficient for TO estimation for our channel model with high XPD.

\subsection{Per-Antenna Timing Offset Estimation\label{subsec:Per-Antenna-Timing-Offset}}

To obtain the per-antenna TO, we first estimate the $NM$ sum-offsets
via finding the maximum correlation peak in the timing correlation
metric. For a high resolution TO estimation, we consider upsampling
at the received signal. Correlating the $Q$-fold upsampled version
of the received signal at the $i$-th receive antenna and the preamble
from the $j$-th transmit antenna, the timing correlation metric at
the $s_{f}$-th sample shift \footnote{Sample shift can be regarded as the sample-level counterpart of \emph{lag}
	in Section \ref{subsec:Preamble-Sequence-Design}.} is given by
\[
\Lambda_{i,j}\left(s_{f}\right)=\left|\sum_{k=0}^{L_{t}-1}a_{j}^{*}\left(k\right)\cdot y_{i}\left(kQ+s_{f}\right)\right|^{2}.
\]
The sum-offset for the $j$-th transmit antenna and the $i$-th receive
antenna, $\tau_{i,j}=\tau_{j}^{tx}+\tau_{i}^{rx}$, is obtained by
selecting the highest peak in the correlation metric, as given by
\[
{\hat{\tau}}_{i,j}=\frac{T_{s}}{T}\cdot\left(\arg\max_{s_{f}=0,...,\lceil2Q\tau_{max}\rceil}\Lambda_{i,j}\left(s_{f}\right)\right).
\]
Then, the per-antenna TO $\boldsymbol{\tau}=\mathbb{R}^{N+M}$ can
be related to the $NM$ sum-offsets by
\begin{equation}
\boldsymbol{\gamma}=\bar{\mathbf{I}}_{NM}\boldsymbol{\tau},\label{eq:sumtoper}
\end{equation}
where $\boldsymbol{\gamma}=\left(\tau_{1,1},...,\tau_{1,M},...,\tau_{N,1},...,\tau_{N,M}\right)^{T}\in\mathbb{R}^{NM}$
and $\bar{\mathbf{I}}_{NM}\in\mathbb{R}^{\left(NM\right)\times\left(N+M\right)}$
is given by
\begin{equation}
\bar{\mathbf{I}}_{NM}=\left[\begin{array}{cc}
\mathbf{I}_{N}\otimes\mathbf{1}_{M} & \mathbf{1}_{N}\otimes\mathbf{I}_{M}\end{array}\right].\label{eq:Adef}
\end{equation}
The column rank of $\bar{\mathbf{I}}_{NM}$ is only $\left(N+M-1\right)$,
thus the transformation (\ref{eq:sumtoper}) is underdetermined and
one additional constraint is required. The reference TO, i.e., $\tau_{1}^{rx}=0$
without loss of generality, provides this additional constraint with
which the per-antenna TO $\boldsymbol{\tau}\in\mathbb{R}^{N+M}$ can
be obtained by solving the following LS problem:
\begin{align}
\hat{\boldsymbol{\tau}}= & \arg\min_{\mathbf{\tau}}\left\Vert \bar{\mathbf{I}}_{NM}\boldsymbol{\tau}-\hat{\boldsymbol{\gamma}}\right\Vert _{2}\nonumber \\
& \ \ \mathrm{s.t.}\ \mathbf{\boldsymbol{\tau}}\left(1\right)=0.\label{eq:LS_timing}
\end{align}

\subsection{Per-Antenna Timing Offset Compensation}

In our system, the TOs are compensated with the corresponding transmit
and receive pulse shaping filters, specifically, with $g\left(t\right)=g^{tx}\left(t\right)*g^{rx}\left(t\right)$,
where $g^{tx}\left(t\right)$ and $g^{rx}\left(t\right)$ represents
the transmit and receive pulse shaping filters, respectively. The
TOs are compensated using $g^{tx}\left(t+\hat{\tau}_{j}^{tx}\right)$
and $g^{rx}\left(t+\hat{\tau}_{i}^{rx}\right)$ at the $j$-th transmit
and $i$-th receive antenna, respectively. Thus, the resulting equivalent
channel after TO compensation will be 
\[
h_{i,j}^{\Delta\boldsymbol{\tau}}\left(t\right)=\left[\tilde{\mathbf{H}}\left(t\right)\right]_{i,j}*g\left(t-\left(\Delta\tau_{j}^{tx}+\Delta\tau_{i}^{rx}\right)T\right),
\]
where $\Delta\tau_{j}^{tx}=\tau_{j}^{tx}-\hat{\tau}_{j}^{tx}$ and
$\Delta\tau_{i}^{rx}=\tau_{i}^{rx}-\hat{\tau}_{i}^{rx}$ represents
the residual TO at the $j$-th transmit and $i$-th receive antenna,
respectively. 

\begin{figure}
	\begin{centering}
		\includegraphics[width=0.5\textwidth]{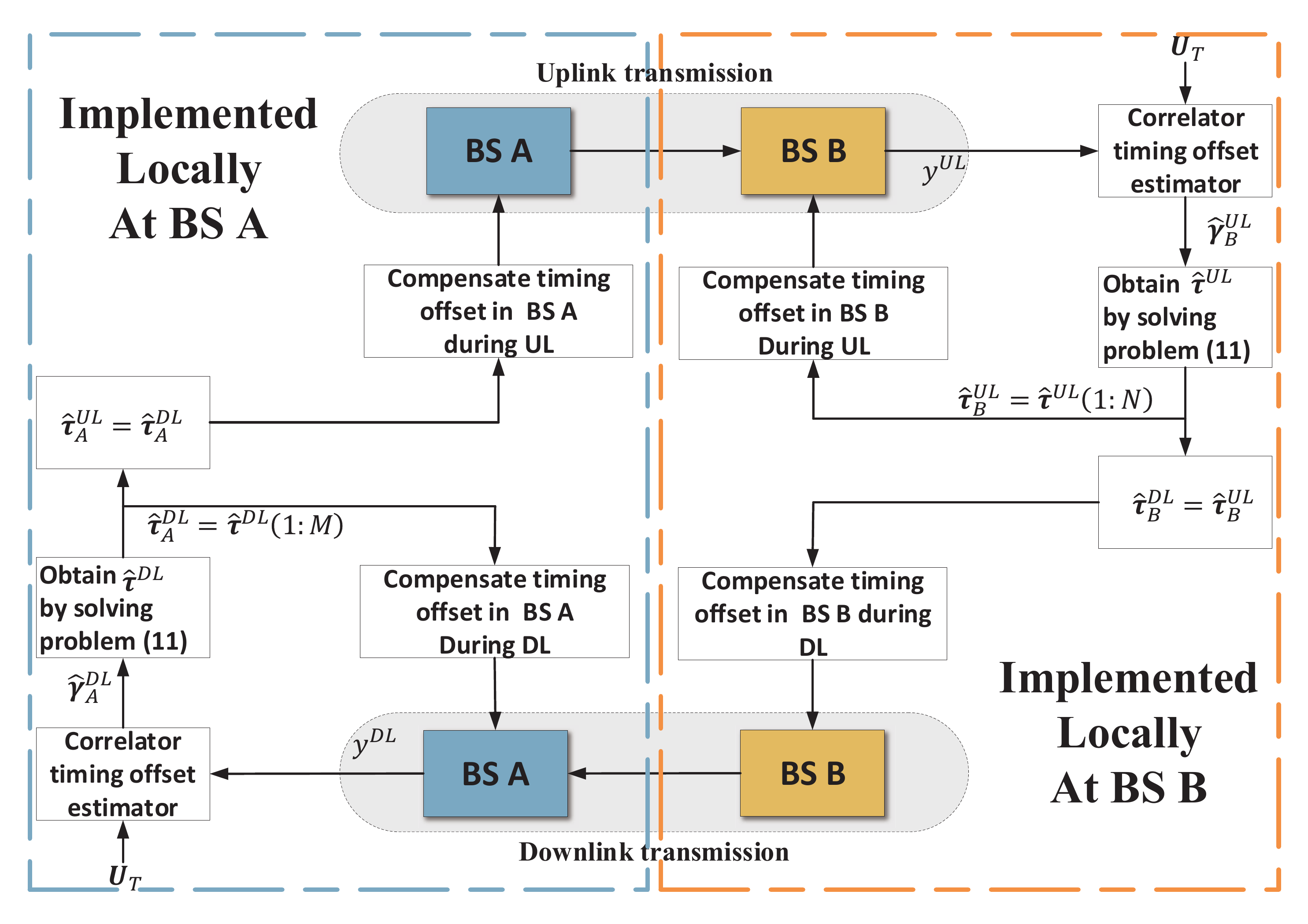}
		\par\end{centering}
	\caption{Illustration of the decentralized per-antenna TO compensation. \label{fig:Per-antenna}}
	
\end{figure}

Since the TOs of BS A (BS B) are identical in the downlink and uplink
transmission, the TO estimated during downlink (uplink) for BS A (BS
B) as a receiver can also be used to compensate the TO during uplink
(downlink) for BS A (BS B) as a transmitter. Thus, the per-antenna
TO compensation can be implemented by only using the local information
in each BS without feedback. For example, consider an FDD system between
BS A (with $M$ antennas) and BS B (with $N$ antennas), as illustrated
in Fig. \ref{fig:Data_path}. The per-antenna TO compensation scheme
is described in Fig. \ref{fig:Per-antenna}. The received samples
in BS B during uplink transmission are used to correlate with the
preamble sequences $\mathbf{U}_{T}$ transmitted by BS A to obtain
the uplink sum-offsets $\hat{\boldsymbol{\gamma}}_{B}^{UL}$, which
is then used to obtain per antenna TO $\hat{\boldsymbol{\tau}}^{UL}$
by solving the LS problem with constraint $\mathbf{\boldsymbol{\tau}}_{B}\left(1\right)=0$
(\ref{eq:LS_timing}). The TOs of the antennas of BS B are then compensated
by the first $N$ elements in $\hat{\boldsymbol{\tau}}^{UL}$, i.e.,
$\hat{\boldsymbol{\tau}}_{B}^{UL}$ in the corresponding transmit
(receive) pulse shaping filters for the downlink (uplink) transmission.
The same compensation procedure is implemented for BS A using downlink
measurements.

\section{Robust MIMO Precoder and Decorrelator Design\label{sec:Robust-MIMO-Precoder}}

In this section, we propose the precoder/decorrelator design for MAI
and ISI suppression. After compensating the TO, there are two physical
impairments that will hinder the performance of the precoder/decorrelator.
One is the residual of the TO estimation, which is within one symbol
time, i.e., $\Delta\mathbf{\tau}_{j}^{tx}\ll T$ and $\Delta\mathbf{\tau}_{i}^{rx}\ll T$,
and these residuals will introduce ISI. Another is the PHN, as illustrated
in (\ref{eq:Symbol-Sig-Model}), which will introduce MAI. For the
convenience of the following discussion, we define $\Delta\boldsymbol{\tau}=\left[\left(\Delta\boldsymbol{\tau}^{rx}\right)^{T},\left(\Delta\boldsymbol{\tau}^{tx}\right)^{T}\right]^{T}=\left[\Delta\tau_{1}^{rx},...,\Delta\tau_{N}^{rx},\Delta\tau_{1}^{tx},...,\Delta\tau_{M}^{tx}\right]^{T}$as
the aggregate residual of the TO estimation and the PHN matrix $\big[\boldsymbol{\Lambda}\left[k\right]\big]{}_{i,j}=\theta_{j}^{tx}\left[k\right]+\theta_{i}^{rx}\left[k\right]$
with $\theta_{j}^{tx}\left[k\right]=\theta_{j}^{tx}\left(kQ\right)$
and $\theta_{i}^{rx}\left[k\right]=\theta_{i}^{rx}\left(kQ\right)$,
as the PHN of the $k$-th symbol at the $j$-th transmit antenna and
the $i$-th receive antenna. Since the operations of downlink and
uplink are similar in this section, we omit the indicator for specifying
the downlink or the uplink for the channel and the precoder/decorrelator.
The channel state information (CSI) is assumed to be static, and the
CSI drift caused by the PHN will be discussed in Section \ref{sec:Phase-Offset-Estimation}.

\subsection{Preliminary for Precoder/decorrelator Design \label{subsec:Channel-Estimation} }

Before presenting the proposed precoder/decorrelator design, the channel
estimation procedure is introduced in this subsection since the CSI
is a fundamental preliminary for the precoder/decorrelator design.
Due to the existence of the residual of the TO estimation, there is
ISI at each symbol transmission. Hence, we need to estimate the CSI
for several taps to support the precoder/decorrelator to suppress
the ISI. We consider the channels within a finite length window with
a $2W+1$ symbol length. Using the discrete expression of the channel
(\ref{eq:equivalent_ch}), the aggregate channel to be estimated is
given by
\begin{equation}
\mathbf{\bar{H}}=\left[\mathbf{H}\left[-W\right],...,\mathbf{H}\left[0\right],...,\mathbf{H}\left[W\right]\right]\in\mathbb{C}^{N\times(2W+1)M},\label{eq:aggc}
\end{equation}
where $\mathbf{H}\left[0\right]\in\mathbb{C}^{N\times M}$ represents
the principal channel corresponding to the current transmitted symbol,
while $\mathbf{H}\left[w\right]$ is defined as the ISI channel with
$w\in\{-W,...,-1,1...,W\}$ corresponding to the adjacent interference
symbols. Each subchannel in $\mathbf{\bar{H}}$ is given by

\begin{equation}
[\mathbf{H}\left[w\right]]_{i,j}=h_{i,j}^{\Delta\tau}\left[w\right]e^{j[\boldsymbol{\Lambda}\left[k^{*}+w\right]]_{i,j}},\forall w\in\{-W,...,0...,W\},\label{eq:ISI_channel}
\end{equation}
where $h_{i,j}^{\Delta\tau}\left[w\right]=h_{i,j}^{\Delta\tau}(wT)$.
$k^{*}$ is the reference symbol; in other words, we assume the principal
channel and ISI channels for each symbol are identical to the principal
channel and ISI channels for the $k^{*}$-th symbol. Since the residual
of the TO estimation is within one symbol time and the channel is
assumed to be static, the input-output relationship at the $k$-th
symbol in the preamble for channel estimation can be expressed as

		\begin{align}
		\mathbf{y}\left(k\right)=\mathbf{\bar{H}}\mathbf{U}_{T}\left(k\right)+\mathbf{\boldsymbol{\nu}}\left[k\right]\label{eq:CEm}
		\end{align}
where $\mathbf{U}_{T}\left(k\right)=\left[\mathbf{a}\left(k+W\right)^{T},...,\mathbf{a}\left(k\right)^{T},...,\mathbf{a}\left(k-W\right)^{T}\right]^{T}\in\mathbb{C}^{(2W+1)M\times1}${\scriptsize{}
}is the reshaped preamble for channel estimation and $\boldsymbol{\mathbf{\mathbf{\nu}}}\left[k\right]=\boldsymbol{\mathbf{\mathbf{\nu}}}\left(kQ\right)$
is the AWGN vector at the \textbf{$k$}-th symbol with variance $\sigma$.
Using $L_{t}$ preamble symbols at the beginning of each frame (except
the initial frame for timing) to estimate the CSI, the received signal
$\mathbf{Y}\in\mathbb{R}^{N\times L_{t}}$ is given by

\begin{align*}
\mathbf{Y} & =\mathbf{\bar{H}}\left[\mathbf{U}_{T}(0),\ldots\mathbf{U}_{T}(L_{t}-1)\right]+\left[\boldsymbol{\nu}\left[0\right],\ldots,\boldsymbol{\nu}\left[L_{t}-1\right]\right]\\
& =\mathbf{\bar{H}}\bar{\mathbf{U}}_{T}+\bar{\boldsymbol{\nu}}.
\end{align*}
Then, the CSI estimation is obtained by the LS solution
\begin{equation}
\hat{\bar{\mathbf{H}}}=\underset{\mathbf{\bar{H}}\in\mathbb{C}^{N\times(2W+1)M}}{\mathrm{argmin}}\left\Vert \mathbf{Y}-\mathbf{\bar{H}}\mathbf{U}_{T}\right\Vert _{F}^{2}=\mathbf{Y}\mathbf{U}_{T}{}^{H}\left(\mathbf{U}_{T}\mathbf{U}_{T}^{H}\right)^{-1},\label{eqCERE}
\end{equation}
where $\hat{\bar{\mathbf{H}}}=\left[\mathbf{\hat{H}}\left[-W\right],...,\mathbf{\hat{H}}\left[0\right],...,\mathbf{\hat{H}}\left[W\right]\right]\in\mathbb{C}^{N\times(2W+1)M}$.

\subsection{Optimization Problem Formulation for Precoder/decorrelator Design}

We use the estimated CSI to optimize the precoder/decorrelator. Consider
a MIMO system, which can support $N_{s}=min\{N,M\}$ parallel transmission
streams. In order to suppress MAI and ISI, we apply a memoryless precoder,
$\text{\ensuremath{\mathbf{F}}\ensuremath{\in\mathbb{C}^{M\times N_{s}}}}$,
at the transmitter, and a memory decorrelator, $\text{\ensuremath{\mathbf{\tilde{W}}}=[\ensuremath{\mathbf{W}(-D)^{H},...\mathbf{W}(D)^{H}]^{H}}\ensuremath{\in\mathbb{C}^{N(2D+1)\times N_{s}}}}$,
at the receiver, where $\mathbf{W}(d)\in\mathbb{C}^{N\times N_{s}}$
is the decorrelator at the $d$-th tap. The received signal in the
$k$-th symbol time at the $m$-th stream can be represented as

\begin{align}
	r_{m}(k) =\mathbf{\tilde{w}}_{m}^{H}\sum_{\substack{w'=-(D+W)}
	}^{(D+W)}\mathbf{\tilde{H}}\left[w'\right]\mathbf{F}\mathbf{s}(k+w')+\mathbf{\tilde{w}}_{m}^{H}\tilde{\boldsymbol{\nu}}\left[k\right],\label{eq:precSig} 
	\end{align}
	where $\mathbf{\tilde{w}}_{m}$  is the $m$-th
	column of  $\ensuremath{\mathbf{\tilde{W}}}$,
	and $\mathbf{\tilde{H}}\left[w'\right]=\left[\begin{array}{c}
	\mathbf{\hat{H}}(w'+D)\\
	\vdots\\
	\mathbf{\hat{H}}(w'-D)
	\end{array}\right]$, $\mathbf{\tilde{\boldsymbol{\nu}}}\left[w'\right]=\left[\begin{array}{c}
\mathbf{\text{\ensuremath{\boldsymbol{\nu}}}}(w'+D)\\
\vdots\\
\mathbf{\mathbf{\text{\ensuremath{\boldsymbol{\nu}}}}}(w'-D)
\end{array}\right]$ is the aggregate channel and noise, respectively, after considering
the memory effect at the decorrelator. $\mathbf{s}(k)=[s_{1}(k),\ldots s_{N_{s}}(k)]$
is the transmitted symbol vector at the $k$-th symbol time in the
data section,\footnote{Note that signal model (\ref{eq:precSig}) can also be applied to
	pilot symbols by substituting $\mathbf{s}(k)$ with $\mathbf{b}(k)$. } and we assume $s_{j}(k)$ $\forall j$ and $\forall k$ are i.i.d
with zero mean unit and variance. 

The existence of both ISI and MAI in Eq.(\ref{eq:precSig}) cannot
be handled by traditional solutions, such as singular value decomposition
(SVD) or naive water filling due to the existence of both ISI and
MAI. Furthermore, considering the practical maximum modulation level
constraint (e.g., we assume that the maximum modulation level that
can be supported in implementation is 4096-QAM), the design of an
optimal precoder/decorrelator is challenging. To achieve a high spectrum
efficiency and incorporate ISI, MAI, and the modulation constraint,
we formulate the following max-sum-rate problem:

\begin{align}
& \underset{\mathbf{F},\mathbf{\ensuremath{\mathbf{\tilde{W}}}}}{{\mathrm{maximize}}}\quad\sum\limits _{m=1}^{N_{s}}\min\{\log_{2}(1+\text{SINR}_{m}),\text{\ensuremath{\varpi}}\}\nonumber \\
& s.t\quad Tr(\mathbf{F^{H}F})\leq P,\label{eq:Precprob}
\end{align}
where $\varpi$ is the capacity under a given QAM modulation level,
$Q_{M}$, with a small error (e.g., symbol error rate (SER) $<10^{-3}$).
$P$ is the total transmit power, and $\text{SINR}_{m}$ is the signal
to interference plus noise ratio (SINR) of the $m$-th stream, which
can be expressed as 
\begin{figure*}
	{\footnotesize
		\begin{align}
		\text{SINR}_{m} & =\frac{\ensuremath{\mathbf{\tilde{w}}_{m}^{H}}\ensuremath{\mathbf{\mathbf{\tilde{H}}}\left[0\right]}\ensuremath{\mathbf{f}_{m}}\mathbf{f}_{m}^{H}\ensuremath{\mathbf{\mathbf{\tilde{H}}}\left[0\right]}^{H}\mathbf{\mathbf{\tilde{w}}}_{m}}{\sum_{\substack{m'\neq m}
			}^{N_{s}}\mathbf{\mathbf{\tilde{w}}}_{m}^{H}\ensuremath{\mathbf{\mathbf{\tilde{H}}}\left[0\right]}\mathbf{f}_{m'}\mathbf{f}_{m'}^{H}\ensuremath{\mathbf{\mathbf{\tilde{H}}}\left[0\right]}\mathbf{\mathbf{\tilde{w}}}_{m}+\sum_{\substack{w'=-(D+W);w'\neq0}
			}^{(D+W)}\mathbf{\tilde{w}}_{m}^{H}\ensuremath{\mathbf{\tilde{H}}\left[w'\right]}\mathbf{F}\mathbf{F}^{H}\mathbf{\mathbf{\tilde{H}}}\left[w'\right]^{H}\mathbf{\mathbf{\tilde{w}}}_{m}+\sigma^{2}\mathbf{\mathbf{\tilde{w}}}_{m}^{H}\mathbf{\mathbf{\tilde{w}}}_{m}}.\label{eq:SINR}
		\end{align}
	}
\end{figure*}

\subsection{Precoder/decorrelator Design via Alternative Optimization }

Problem (\ref{eq:Precprob}) is non-convex and non-smooth, hence is
very challenging to solve. Inspired by \cite{shi2011iteratively},
we introduce an auxiliary variable $\boldsymbol{\Gamma}\in\mathbb{C}^{N_{s}\times N_{s}}$
and use an alternative optimization scheme to find a simple solution
to problem (\ref{eq:Precprob}). Using the variable $\boldsymbol{\Gamma}$,
problem (\ref{eq:Precprob}) can be well approximated by 

\begin{align}
& \underset{\mathbf{F},\mathbf{\ensuremath{\mathbf{\tilde{W}}}},\boldsymbol{\Gamma}}{{\mathrm{minimize}}}\quad Tr\big(\boldsymbol{\Gamma}\sum\limits _{m=1}^{N_{s}}\mathbf{I}_{m}^{T}\mathbf{E}(\mathbf{W},\mathbf{F})\mathbf{I}_{m}\big)-\log\det(\boldsymbol{\Gamma})\nonumber \\
& s.t\quad Tr(\mathbf{F^{H}F})\leq P,\label{eq:Precprobapp}\\
& \quad\quad[\boldsymbol{\Gamma}]_{m,m}\leq2^{\text{\ensuremath{\varpi}}},\forall m,\nonumber 
\end{align}
where $\mathbf{I}_{m}$ is the diagonal matrix with diagonal elements
as the elements in the $m$-th column of the identity matrix, and
\begin{align}
\mathbf{E}(\mathbf{\ensuremath{\mathbf{\tilde{W}}}},\mathbf{F}) & =(\mathbf{\tilde{W}}^{H}\mathbf{\tilde{H}}\left[0\right]\mathbf{F}-\mathbf{I})(\mathbf{F}^{H}\mathbf{\tilde{H}}\left[0\right]^{H}\mathbf{\tilde{W}}-\mathbf{I})\nonumber \\
+ & \mathbf{\tilde{W}}^{H}\big(\sum_{\substack{w'=-(D+W);w'\neq0}
}^{(D+W)}\mathbf{\mathbf{\tilde{H}}}[w']\mathbf{F}\mathbf{F}{}^{H}\mathbf{\mathbf{\tilde{H}}}[w']^{H}+\sigma^{2}\mathbf{I}\big)\mathbf{\ensuremath{\mathbf{\tilde{W}}}}.\label{eq:calE}
\end{align}
The details of this approximation can be found in Appendix \ref{subsec:PRE}. 

We then present the detailed update rules using the alternative optimization
method for problem (\ref{eq:Precprobapp}). At the $t_{iter}$-th
iteration of the optimization, we alternatively update $\mathbf{\ensuremath{\mathbf{\tilde{W}}}},$
$\boldsymbol{\Gamma}$ and $\mathbf{F}$ by the following 3 steps:
\begin{itemize}
	\item Step 1: Update $\ensuremath{\mathbf{\tilde{W}}}^{(t_{iter})}$ given
	$\boldsymbol{\Gamma}^{(t_{iter}-1)}$ and $\mathbf{F}^{(t_{iter}-1)}$
	by 
	\begin{equation}
	\ensuremath{\mathbf{\tilde{W}}}^{(t_{iter})}=\mathbf{B}^{-1}\mathbf{\mathbf{\tilde{H}}}[0]\mathbf{F}^{(t_{iter}-1)},\label{eq:upW-1}
	\end{equation}
	where
	\begin{align}
	\mathbf{B} & =\mathbf{\mathbf{\tilde{H}}}[0]\mathbf{F}^{(t_{iter}-1)}(\mathbf{F}^{(t_{iter}-1)}){}^{H}\mathbf{\mathbf{\tilde{H}}}^{H}[0]\nonumber \\
	& +\sum_{\substack{w'=-(D+W);w'\neq0}
	}^{(D+W)}\mathbf{\mathbf{\tilde{H}}}[w']\mathbf{F}^{(t_{iter}-1)}(\mathbf{F}^{(t_{iter}-1)}){}^{H}\mathbf{\mathbf{\tilde{H}}}[w']^{H}\label{eq:Wb-1}\\
	& +\sigma^{2}\mathbf{I}\text{.}\nonumber 
	\end{align}
	\item Step 2: Update $\boldsymbol{\Gamma}^{(t_{iter})}$ given $\ensuremath{\mathbf{\tilde{W}}}^{(t_{iter})}$
	and $\mathbf{F}^{(t_{iter}-1)}$ by 
	\begin{equation}
	\boldsymbol{\Gamma}^{(t_{iter})}=\min\{[\mathbf{E}(\ensuremath{\mathbf{\tilde{W}}}^{(t_{iter})},\mathbf{F}^{(t_{iter}-1)})]_{m,m}^{-1},2^{\text{\ensuremath{\varpi}}}\}.\label{eq:upga}
	\end{equation}
	\item Step 3: Update $\mathbf{F}^{(t_{iter})}$ given $\ensuremath{\mathbf{\tilde{W}}}^{(t_{iter})}$
	and $\boldsymbol{\Gamma}^{(t_{iter})}$ by solving\\
	\begin{align}
	& \underset{\mathbf{F}}{{\mathrm{minimize}}}\quad Tr\big(\boldsymbol{\Gamma}^{(t_{iter})}\sum\limits _{m=1}^{N_{s}}\mathbf{I}_{m}^{T}\mathbf{E}(\ensuremath{\mathbf{\tilde{W}}}^{(t_{iter})},\mathbf{F})\mathbf{I}_{m}\big)\nonumber \\
	& s.t\quad Tr(\mathbf{F^{H}F})\leq P.\label{eq:SubforF-1}
	\end{align}
	Problem (\ref{eq:SubforF-1}) is convex, which can be efficiently solved by any solver for convex problems, e.g., CVX Matlab package \cite{cvx}. 
\end{itemize}
These 3 update rules are obtained by the property that problem (\ref{eq:Precprobapp})
becomes convex in terms of any individual variable in $\mathbf{F},\mathbf{\ensuremath{\mathbf{\tilde{W}}}},\boldsymbol{\Gamma}$
when fixing the other two. This property guarantees the sequence generated
by the above alternative optimization converges to a stationary point
of problem (\ref{eq:Precprobapp}). To summarize, the precoder/decorrelator
can be obtained via Algorithm \ref{alg1-1}.

\begin{algorithm}
	\caption{\label{alg1-1}Alternative optimization algorithm for precoder/decorrelator
		design}
	
	\begin{algorithmic}[1]
		
		\STATE\textbf{Input:} Estimated CSI $\hat{\bar{\mathbf{H}}}$, total
		power $P$ and noise variance $\sigma^{2}.$
		\begin{raggedright}
			\STATE\textbf{Output:} $\hat{\mathbf{F}}$,$\hat{\mathbf{W}}$.
			\par\end{raggedright}
		\STATE\textbf{Initialize:} Construct $\mathbf{\tilde{H}}\left[w'\right],w'\in\{-(D+W),\ldots,(D+W)\}$
		by $\mathbf{\hat{H}}[w]$. $t_{iter}=0$, any$\text{\ensuremath{\mathbf{F}^{(0)}} }$
		satisfy the power constraint, $\boldsymbol{\Gamma}^{(0)}=diag(2^{\text{\ensuremath{\varpi}}},\ldots2^{\text{\ensuremath{\varpi}}})$.
		
		\WHILE{not converge} 
		
		\STATE $t_{iter}=t_{iter}+1$
		
		\STATE Update $\mathbf{\mathbf{\tilde{W}}}^{(t_{iter})}$ according
		to Eq.(\ref{eq:upW-1}) and Eq.(\ref{eq:Wb-1}) with $\mathbf{F}^{(t_{iter}-1)}$,
		$\boldsymbol{\Gamma}^{(t_{iter}-1)}$.
		
		\STATE Update $\boldsymbol{\Gamma}^{(t_{iter})}$ according to Eq.(\ref{eq:upga})
		and Eq.(\ref{eq:calE}) with $\mathbf{F}^{(t_{iter}-1)}$, $\mathbf{W}^{(t_{iter})}$.
		
		\STATE Update $\mathbf{F}^{(t_{iter})}$ by solving problem (\ref{eq:SubforF-1})
		with $\mathbf{\mathbf{\tilde{W}}}^{(t_{iter})}$, $\boldsymbol{\Gamma}^{(t_{iter})}$
		by any solver for convex problems.
		
		\ENDWHILE
		
		\STATE$\hat{\mathbf{F}}=\mathbf{F}^{(t_{iter})}$, $\hat{\mathbf{W}}=\mathbf{\mathbf{\tilde{W}}}^{(t_{iter})}$.
		
	\end{algorithmic}
\end{algorithm}

\section{Phase Noise Estimation and Compensation\label{sec:Phase-Offset-Estimation}}

After the precoder and decorrelator, the MAI and ISI is suppressed
to enable high spectral efficiency transmission. However, the phase
of the effective channel changes after the channel estimation stage
due to the drifting of the PHN, and thus there will be accumulating
MAI caused by the coherence loss in the precoder and decorrelator.
As a result, the PHN needs to be tracked and compensated in the data
transmission stage, which is the main target of this section.

\subsection{Per-Antenna PHN Estimation based on Pilots\label{subsec:Per-Antenna-PHN-Estimation}}

From equation (\ref{eq:ISI_channel}) and (\ref{eq:CEm}), the channel
estimation stage absorbs the initial PHN into the channel phase response.
But as the PHN continues to drift, the effect of coherency loss will
become dominant and introduce more MAI. To reset the phase error over
one frame, we utilize the $N_{sf}-1$ pilots inserted in a frame after
the preamble to estimate the PHN increment from the preamble to each
subframe. However, the presence of TO error and multipath creates
non-negligible ISI, which deteriorates the PHN estimation quality.
Though we can accurately estimate the sum-PHN in each ISI channel
matrix as in the channel estimation stage, it requires a long pilot
sequence to achieve high accuracy, which introduces significant overhead.
In this section, we exploit the interference suppression capability
of the precoder/decorrelator to improve the estimation accuracy.

Since the PHN varies very slowly within each pilot sequence, we assume
that the PHN is constant during pilot sequence transmission in each
subframe. We define the \emph{accumulated PHN increment} from the
preamble to the $q$-th subframe as $\phi_{j}^{tx}\left[q\right]=\theta_{j}^{tx}\left[q(L_{d}+L_{p})\right]-\theta_{j}^{tx}\left[k^{*}\right]$
and $\phi_{i}^{rx}\left[q\right]=\theta_{i}^{rx}\left[q(L_{d}+L_{p})\right]-\theta_{i}^{rx}\left[k^{*}\right]$
for the $j$-th transmit antenna and the $i$-th receive antenna,
where $k^{*}$ is the reference symbol we picked in the preamble for
channel estimation (as elaborated in Eq.(\ref{eq:ISI_channel})).
The collection of the\emph{ accumulated PHN increments} at the $q$-th
subframe from all $M$ transmit antennas and from all $N$ receive
antennas are denoted by
$\boldsymbol{\phi}^{tx}\left[q\right]\triangleq\left[\phi_{1}^{tx}\left[q\right],...,\phi_{M}^{tx}\left[q\right]\right]^{T}$
and $\boldsymbol{\phi}^{rx}\left[q\right]\triangleq\left[\phi_{1}^{rx}\left[q\right],...,\phi_{N}^{rx}\left[q\right]\right]^{T}$.
We assume that after the processing by the precoder/decorrelator,
the ISI in the pilot transmissions is canceled, hence the received
symbols at the $\left(q+1\right)$-th pilot is given by
\begin{align*}
& \mathbf{R}\left[q+1\right]\\
= & \sum_{d=-D}^{D}\mathbf{W}\left[d\right]^{H}\underbrace{\mathbf{D}_{\Delta\boldsymbol{\phi}}^{rx}\left[q\right]\mathbf{H}\left[-d\right]\mathbf{D}_{\Delta\boldsymbol{\phi}}^{tx}\left[q\right]}_{\mathbf{H}_{q+1}\left[d\right]}\mathbf{F}\mathbf{U}_{P}+\mathbf{V}\left[q+1\right]\\
= & \sum_{d=-D}^{D}\mathbf{W}\left[d\right]^{H}\mathbf{H}_{q+1}\left[d\right]\mathbf{X}+\mathbf{V}\left[q+1\right],
\end{align*}
where 
\begin{itemize}
	\item $\mathbf{D}_{\Delta\mathbf{\phi}}^{rx}\left[q\right]\triangleq\mathrm{diag}\left(e^{j\Delta\boldsymbol{\phi}^{rx}\left[q\right]}\right)$
	is an $M\times M$ diagonal matrix, where $\Delta\boldsymbol{\phi}^{rx}\left[q\right]\triangleq\boldsymbol{\phi}^{rx}\left[q+1\right]-\hat{\boldsymbol{\phi}}^{rx}\left[q\right]$
	collects the PHN increments at the receiver;
	\item $\mathbf{D}_{\Delta\boldsymbol{\phi}}^{tx}\left[q\right]\triangleq\mathrm{diag}\left(e^{j\Delta\boldsymbol{\phi}^{tx}\left[q\right]}\right)$
	is an $N\times N$ diagonal matrix, where $\Delta\boldsymbol{\phi}^{tx}\left[q\right]\triangleq\boldsymbol{\phi}^{tx}\left[q+1\right]-\hat{\boldsymbol{\phi}}^{tx}\left[q\right]$
	collects the PHN increments at the transmitter;
	\item $\mathbf{H}_{q+1}\left[d\right]\triangleq\mathbf{D}_{\Delta\boldsymbol{\phi}}^{rx}\left[q\right]\mathbf{H}\left[-d\right]\mathbf{D}_{\Delta\boldsymbol{\phi}}^{tx}\left[q\right]$
	is the channel corresponding to the desired pilot signal for tap $d$
	at the beginning of the $\left(q+1\right)$-th subframe;
	\item $\mathbf{X}=\mathbf{F}\mathbf{U}_{P}$ denotes the transmitted signal
	after applying the precoder $\mathbf{F}$ to the pilots $\mathbf{U}_{P}$
	and
	\item $\mathbf{V}\left[q+1\right]\in\mathbb{C}^{N_{s}\times L_{p}}$ contains
	the AWGN noise.
\end{itemize}
To estimate of the \emph{PHN increments }at\emph{ $\mathbf{R}\left[q+1\right]$},
we use first order Taylor expansion $e^{j\Delta\phi}\approx1+j\Delta\phi$
to approximate $\mathbf{D}_{\Delta\boldsymbol{\phi}}^{rx}\left[q\right]$
and $\mathbf{D}_{\Delta\boldsymbol{\phi}}^{tx}\left[q\right]$ by
$\mathbf{I}+\mathrm{diag}(j\Delta\boldsymbol{\phi}^{tx}\left[q\right])$
and $\mathbf{I}+\mathrm{diag}(j\Delta\boldsymbol{\phi}^{rx}\left[q\right])$.
We also define $\boldsymbol{\Omega}_{m}\left(k\right)\in\mathbb{R}^{NM\times NM}$,
with $\left[\boldsymbol{\Omega}_{m}^{d}\left(k\right)\right]_{i,j}=[\mathbf{W}\left(d\right)^{H}]_{i,m}\mathbf{H}\left[-d\right]_{i,j}\left[\mathbf{X}\right]_{j,k}$.
The collection of the \emph{PHN increments $\Delta\boldsymbol{\phi}\left[q\right]\triangleq\left[\Delta\boldsymbol{\phi}^{rx}\left[q\right]^{T},\Delta\boldsymbol{\phi}^{tx}\left[q\right]^{T}\right]^{T}\in\mathbb{R}^{M+N}$
}can be obtained by solving the following LS problem:
\begin{align}
\Delta\hat{\boldsymbol{\phi}}\left[q\right] & =\arg\min_{\Delta\boldsymbol{\phi}\left[q\right]}\left\Vert \boldsymbol{\Xi}\Delta\boldsymbol{\phi}\left[q\right]-\left(\mathrm{vec}(\mathbf{R}\left[q+1\right])-\boldsymbol{\zeta}\right)\right\Vert _{2}^{2}\nonumber \\
\mathrm{s.t.} & \Delta\boldsymbol{\phi}_{1}\left[q\right]=0,\label{eq:PHN_LS}
\end{align}
where 
\begin{align*}
\boldsymbol{\Xi} & =\left[\boldsymbol{\xi}_{1}^{T}\left(1\right),...,\boldsymbol{\xi}_{N_{s}}^{T}\left(1\right),\ldots,\boldsymbol{\xi}_{1}^{T}\left(L_{p}\right),\ldots,\boldsymbol{\xi}_{N_{s}}^{T}\left(L_{p}\right)\right]
\end{align*}
with 
\begin{equation}
\boldsymbol{\xi}_{m}^{T}\left(k\right)=\mathbf{1}_{MN\left(2D+1\right)}^{T}\left[\begin{array}{c}
\mathrm{diag}\left\{ \mathrm{vec}\left[j\left(\boldsymbol{\Omega}_{m}^{-D}\left(k\right)\right)^{T}\right]\right\} \\
\vdots\\
\mathrm{diag}\left\{ \mathrm{vec}\left[j\left(\boldsymbol{\Omega}_{m}^{D}\left(k\right)\right)^{T}\right]\right\} 
\end{array}\right]\bar{\mathbf{I}}_{NM}\label{eq:vector-define}
\end{equation}
$\forall m=1,\ldots N_{s},k=1,\ldots L_{p}$, and we defined $\boldsymbol{\zeta}=\Bigg[\mathbf{1}_{MN\left(2D+1\right)}^{T}\boldsymbol{\eta}_{1}\left(1\right),\ldots,\mathbf{1}_{MN\left(2D+1\right)}^{T}\boldsymbol{\eta}_{N_{s}}\left(L_{p}\right)\Bigg]^{T}$
with
\begin{equation}
\boldsymbol{\eta}_{m}\left(k\right)=\left[\mathrm{vec}\left[\boldsymbol{\Omega}_{m}^{-D}\left(k\right)\right]^{T},...,\mathrm{vec}\left[\boldsymbol{\Omega}_{m}^{D}\left(k\right)\right]^{T}\right]^{T}\label{eq:eta-define}
\end{equation}
$\forall m=1,\ldots N_{s},k=1,\ldots L_{p}.$ Problem (\ref{eq:PHN_LS})
is derived by expanding $\mathbf{R}\left[q+1\right]$ elementwisely,
and the detailed derivation can be found in Appendix \ref{subsec:DerivationPHNLS}.

The PHN at $\left(q+1\right)$-th pilot is then updated by
\[
\hat{\boldsymbol{\phi}}\left[q+1\right]=\hat{\boldsymbol{\phi}}\left[q\right]+\Delta\hat{\boldsymbol{\phi}}\left[q\right]\ \mathrm{for}\ q=0,...,N_{sf}-1,
\]
where $\hat{\boldsymbol{\phi}}\left[0\right]=0$ as we
assume PHN at the preamble is absorbed to the channel and is perfectly
estimated during the channel estimation stage.

\subsection{Per-Antenna PHN Compensation}

We focus on the per-antenna PHN compensation in this subsection. Traditionally,
the PHN information obtained at the receiver is feedback to the transmitter
for PHN compensation. However, per-symbol feedback of PHN estimates
will incur too much overhead. Fortunately, the transmitter and receiver
share the same oscillator in an FDD system so that the PHN is identical
in the uplink and downlink at the same site, which enables local compensation
of the PHN and avoids the overhead for PHN feedback. Consider the
transmission datapath in Fig. \ref{fig:Data_path}, the PHN estimated
during downlink (uplink) for BS A (BS B) as a receiver can also be
used to compensate the PHN during uplink (downlink) for BS A (BS B)
as a transmitter. Thus, PHN compensation can be implemented by using
local information only without feedback.

However, unlike the TO compensation, where the reference $\tau_{1}$
is usually known a priori, $\Delta\phi\left[q\right]_{1}$ in (\ref{eq:PHN_LS})
is not available in advance to both BS A and BS B. Thus, there will
be a \emph{common PHN error} across all the antennas at the receiver.
If we directly use the result $\Delta\hat{\boldsymbol{\phi}}\left[q\right]$
to compensate the PHN and the common PHN error is identical to the
PHN of the first link, i.e., the sum-PHN at the first receive antenna
and first transmit antenna. For example, let 
\[
\Delta\hat{\boldsymbol{\phi}}^{UL}\left[q\right]=\left[\left(\Delta\hat{\boldsymbol{\phi}}_{B}^{UL}\left[q\right]\right)^{T},\left(\Delta\hat{\boldsymbol{\phi}}_{A}^{UL}\left[q\right]\right)^{T}\right]^{T}
\]
and 
\[
\Delta\hat{\boldsymbol{\phi}}^{DL}\left[q\right]=\left[\left(\Delta\hat{\boldsymbol{\phi}}_{A}^{DL}\left[q\right]\right)^{T},\left(\Delta\hat{\boldsymbol{\phi}}_{B}^{DL}\left[q\right]\right)^{T}\right]^{T}
\]
be the estimated $\left(M+N\right)$ PHN vector obtained by solving
problem (\ref{eq:PHN_LS}) in BS B during uplink and BS A during downlink,
respectively. When BS A and BS B compensate the PHN using local information
$\Delta\hat{\boldsymbol{\phi}}_{A}^{DL}\left[q\right]$
and $\Delta\hat{\boldsymbol{\phi}}_{B}^{UL}\left[q\right]$
respectively, the common PHN error at BS A and BS B can be calculated
locally by $\hat{\phi}_{A,CM}^{DL}\left[q+1\right]=\Delta\hat{\phi}_{A,1}^{DL}\left[q\right]+\Delta\hat{\phi}_{B,1}^{DL}\left[q\right]$
and $\hat{\phi}_{B,CM}^{UL}\left[q+1\right]=\Delta\hat{\phi}_{B,1}^{UL}\left[q\right]+\Delta\hat{\phi}_{A,1}^{UL}\left[q\right]$,
where $\Delta\hat{\phi}_{A,i}^{DL}\left[q\right]$ denotes the $i$-th
element of $\Delta\hat{\boldsymbol{\phi}}_{A}^{DL}\left[q\right]$. 

\begin{figure}
	\begin{centering}
		\includegraphics[width=0.45\textwidth]{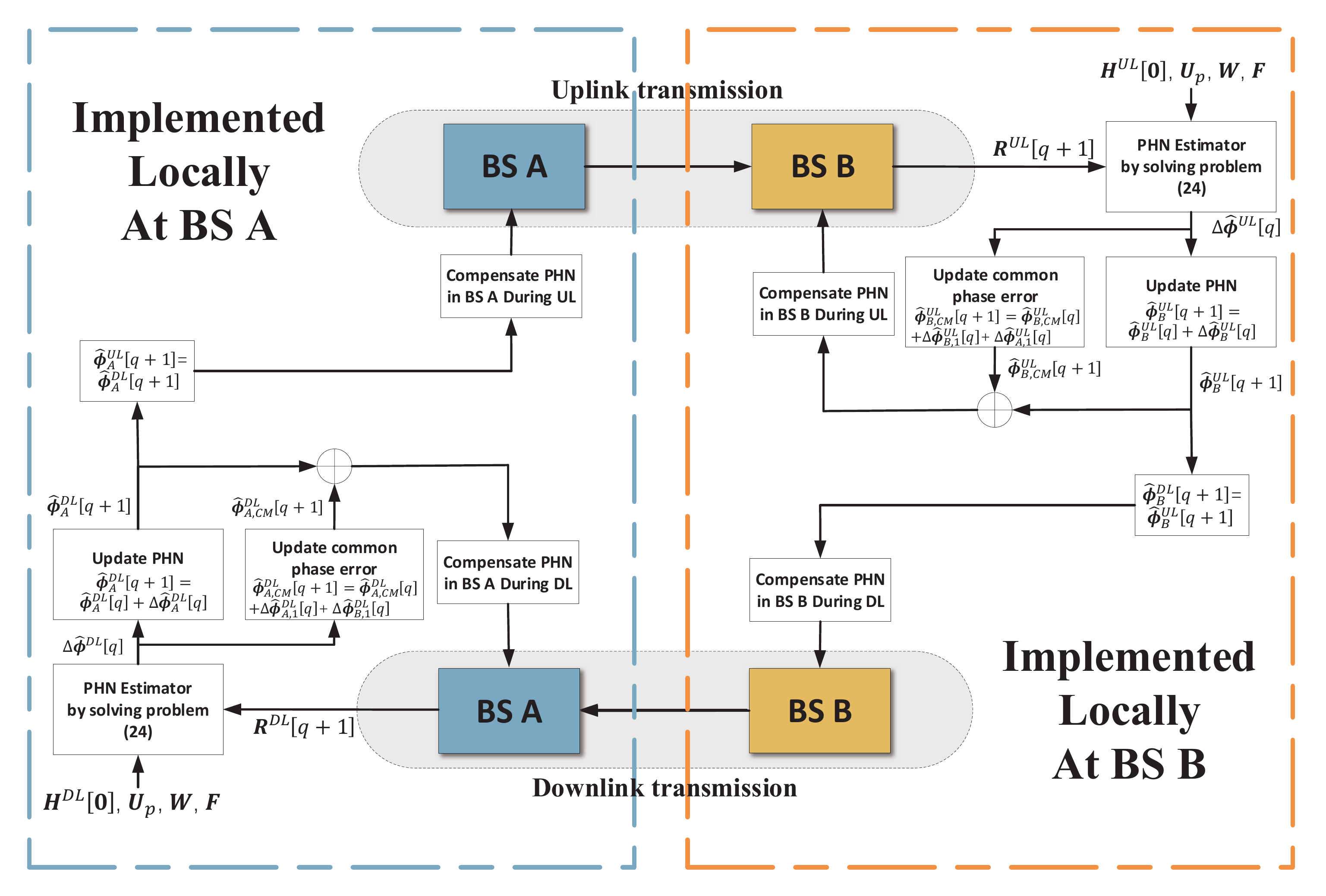}
		\par\end{centering}
	\caption{Per-antenna PHN compensation scheme.\label{fig:Per-antenna-PHN}}
	
\end{figure}

The per-antenna PHN compensation scheme is described in Fig. \ref{fig:Per-antenna-PHN}.
The received symbols $\mathbf{R}^{UL}\left[q+1\right]$ in BS B during
uplink transmission are used to compute the estimation for PHN increment
$\Delta\hat{\boldsymbol{\phi}}^{UL}\left[q\right]$ by
solving problem (\ref{eq:PHN_LS}), which is then used to obtain per
antenna PHN accumulation $\hat{\boldsymbol{\phi}}_{B}^{UL}\left[q+1\right]$
and common phase error accumulation $\hat{\phi}_{B,CM}^{UL}\left[q+1\right]$
at BS B. The PHN of the antennas of BS B are then compensated by $\hat{\boldsymbol{\phi}}_{B}^{UL}\left[q+1\right]$
in the downlink, while they are compensated by $\hat{\boldsymbol{\phi}}_{B}^{UL}\left[q+1\right]+\hat{\phi}_{B,CM}^{UL}\left[q+1\right]$
during the uplink. The same compensation procedure is implemented
for BS A using the downlink measurements.

\subsection{Per-Antenna PHN Tracking based on Decision Feedback}

To track the PHN parameter of each antenna in a subframe, we proposed
a decision-feedback (DFB) estimator to support PHN estimation during
data transmission between consecutive pilots.

The $L_{d}$ symbols in a subframe is split to $N_{b}$ data blocks
where each data block contains $L_{b}$ symbols such that $L_{d}=N_{b}L_{b}$.
In the practical scenario of interest, the PHN is varying slowly compared
to symbol time, the PHNs during each data block are assumed to be
constant and we denote the PHN of the $p$-th data block in the $q$-th
subframe by $\boldsymbol{\phi}^{tx}\left[p;q\right]$
and $\boldsymbol{\phi}^{rx}\left[p;q\right]$ for the
transmitter and receiver, respectively, and define $\boldsymbol{\phi}\left[p;q\right]=\left[\boldsymbol{\phi}^{rx}\left[p;q\right]^{T},\boldsymbol{\phi}^{tx}\left[p;q\right]^{T}\right]^{T}$.
Therefore, in the DFB case, the $L_{b}$ previously detected data
symbols $\bar{\mathbf{S}}\left[p;q\right]$ in the $p$-th data block
of the $q$-th subframe are used to estimate the PHN parameters $\boldsymbol{\phi}\left[p;q\right]$.
For simplicity, we assume perfect decision-feedback during data transmission.

Based on the above assumptions, the estimate of PHN increment $\Delta\boldsymbol{\phi}\left[p;q\right]=\left[\Delta\boldsymbol{\phi}^{rx}\left[p;q\right]^{T},\Delta\boldsymbol{\phi}^{tx}\left[p;q\right]^{T}\right]^{T}$
from the $p$-th to the $\left(p+1\right)$-th data block is obtained
by solving{\footnotesize{}
	\begin{align}
	\Delta\hat{\boldsymbol{\phi}}^{DFB}\left[p;q\right] & =\arg\min_{\Delta\boldsymbol{\phi}}\left\Vert \mathbf{R}\left[p+1;q\right]-\right.\nonumber \\
	& \left.\sum_{d=-D}^{D}\mathbf{W}\left(d\right)^{H}\mathbf{D}_{\Delta\boldsymbol{\phi}}^{rx}\mathbf{H}\left[-d\right]\mathbf{D}_{\Delta\boldsymbol{\phi}}^{tx}\mathbf{F}\bar{\mathbf{X}}\left[p+1;q\right]\right\Vert ^{2},\label{eq:LS-norm-DFB}
	\end{align}
}where $\mathbf{R}\left[p+1;q\right]$ is the received symbols of
the $\left(p+1\right)$-th data block in the $q$-th subframe, and
$\bar{\mathbf{X}}\left[p+1;q\right]=\mathbf{F}\bar{\mathbf{S}}\left[p+1;q\right]$.
The problem can also be approximated by the Taylor expansion on small
PHN increments between data blocks and solved using the same technique
introduced in pilot PHN estimation as specified in Section \ref{subsec:Per-Antenna-PHN-Estimation}.
The estimation for accumulated PHN via DFB at the $\left(p+1\right)$-th
data block is then updated by
\[
\hat{\boldsymbol{\phi}}^{DFB}\left[p+1;q\right]=\hat{\boldsymbol{\phi}}\left[p;q\right]+\Delta\hat{\boldsymbol{\phi}}^{DFB}\left[p;q\right].
\]
However, the DFB estimator may be prone to errors in the detection
of symbols. Instead, we use moving window averaging to track the evolution
of the PHN, i.e., the estimated PHN of $\boldsymbol{\phi}\left[p+1;q\right]$
is given by
\begin{equation}
\hat{\boldsymbol{\phi}}\left[p+1;q\right]=\left(1-\alpha\right)\hat{\boldsymbol{\phi}}\left[p;q\right]+\alpha\hat{\boldsymbol{\phi}}^{DFB}\left[p+1;q\right],\label{eq:Moving-avg}
\end{equation}
where $0\leq\alpha\leq1$, $p=0,1,...,N_{b}$, and $\hat{\boldsymbol{\phi}}\left[0;q\right]=\hat{\boldsymbol{\phi}}\left[q\right]$
takes the pilot-aided PHN estimation result. In (\ref{eq:Moving-avg}),
$\hat{\boldsymbol{\phi}}\left[p;q\right]$ is the estimated
PHN from the history, while $\hat{\boldsymbol{\phi}}^{DFB}\left[p+1;q\right]$
is the new PHN estimate from the DFB in the $\left(p+1\right)$th
data block. By adjusting $\alpha$, we strike a balance between the
history and the innovation from the DFB.

\section{Simulations and Discussions\label{sec:Simulations-and-Discussions}}

In this section, the performance of the proposed massive MIMO mmWave
backhaul system is evaluated. In the following results, we consider
multiple practical cases where $23$ GHz carrier is used and $8\times8$
flat-panel dual-polarized MIMO antennas arrays are equipped at the
transmitter and receiver to transmit data at a distance $D=3$ km,
as illustrated in Fig. \ref{fig:Illustration-of-antenna}. Within
a single the antenna array, antenna element spacing is $d1=\frac{\lambda}{2}$,
and the XPD between H mode and V mode in each dual polar element varies
from $0$ dB to $30$ dB. All the transmitted signals have been normalized
to have unit average power, with symbol duration $T=40$ ns. The propagation
delay associated with the interpath is $\tau_{d}=6.3$ ns and the
notch depth $\rho$ in the reflected path varies from $10$ dB to
$15$ dB. The variance of the AWGN, is set to $\sigma^{2}=\frac{1}{SNR}$
with SNR=$47$ dB. The maximum TOs is $\tau_{max}\cdot T=200\mathrm{ns}$
($5$ symbols time). The variance of the PHN increment process over
one symbol duration is $\sigma_{\Delta}^{2}=10^{-6}$$\mathrm{rad}^{2}$
and we set $\alpha=0.1$ for PHN tracking. The frame structure parameters
are chosen as $L_{t}=256$, $L_{p}=64$, $L_{d}=1280$ and $N_{sf}=100$.
The number of taps for the ISI channel and the memory decorrelator
are respectively, $W=3$ and $D=3$. In the simulation results, we
will first demonstrate the efficiency of the proposed methods in each
component of this system, i.e., TO estimation and compensation, precoder
and decorrelator, PHN estimation and compensation. Finally, we will
show the end-to-end simulation results, which incorporate all the
components proposed in this work.

\subsection{Performance of Individual Design}

\begin{figure}
	\begin{centering}
		\includegraphics[scale=0.55]{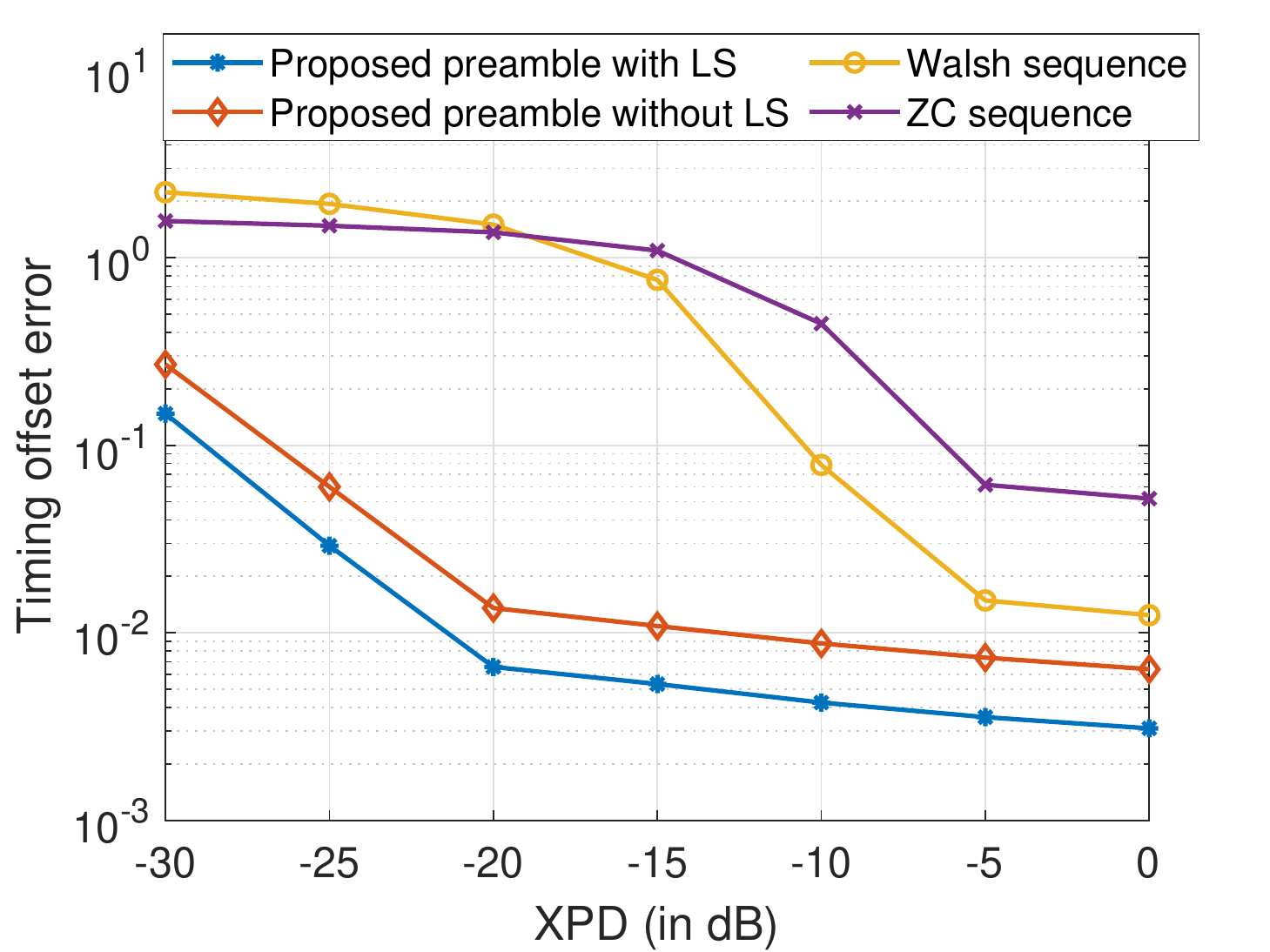}\caption{TO estimation error using the proposed preamble sequence,the ZC sequence,
			and theWalsh sequence.\label{fig:Timing-estimation-performance:}}
		\par\end{centering}
\end{figure}

Figure. \ref{fig:Timing-estimation-performance:} shows the performance
comparison of the LS-based TO estimator with the proposed preamble
sequence, traditional ZC sequence, and Walsh sequence of same length
$L_{t}=256$. The performance metric for TO estimation is the square
root of the MSE of the effective sum-offset in each MIMO link, which
is defined as $\sqrt{\mathbb{E}\left[\frac{||\boldsymbol{\hat{\gamma}}-\boldsymbol{\gamma}||_{2}^{2}}{N^{2}}\right]}$,
where $\boldsymbol{\gamma}$ is the sum-offsets as defined in Section
\ref{sec:Joint-Timing-Estimation}, and the expectation is taken over
500 realizations. \emph{ZC sequence }and\emph{ Walsh sequence} act
as the baselines to estimate the sum-offsets via the LS-based method.
\emph{Proposed preamble without LS }is another baseline that only
adopts the proposed preamble to directly estimate the sum-offsets
without the LS procedure in (\ref{eq:LS_timing}). The\emph{ proposed
	preamble with LS} is the proposed TOs estimation method, in which
$\hat{\boldsymbol{\gamma}}\in\mathcal{\mathbb{R}}^{N^{2}\times1}$
is reconstructed from $\hat{\boldsymbol{\tau}}\in\mathbb{R}^{2N\times1}$
by equation (\ref{eq:LS_timing}). As shown in Fig. \ref{fig:Timing-estimation-performance:},
the proposed method shows superior performance over the other baselines
with XPD ranging from $0$ dB to $30$ dB, and the proposed preamble
is more robust to higher XPD. The proposed preamble only shows a significant
performance decay when XPD $\geq20$ dB. However, the performance
of the ZC sequence and Walsh sequence degrades significantly for XPD
$\geq5$ dB. The LS procedure gives us more than a 2 times higher
accuracy in terms of square root MSE for the proposed preamble.

\begin{figure}
	\centering{}\includegraphics[scale=0.45]{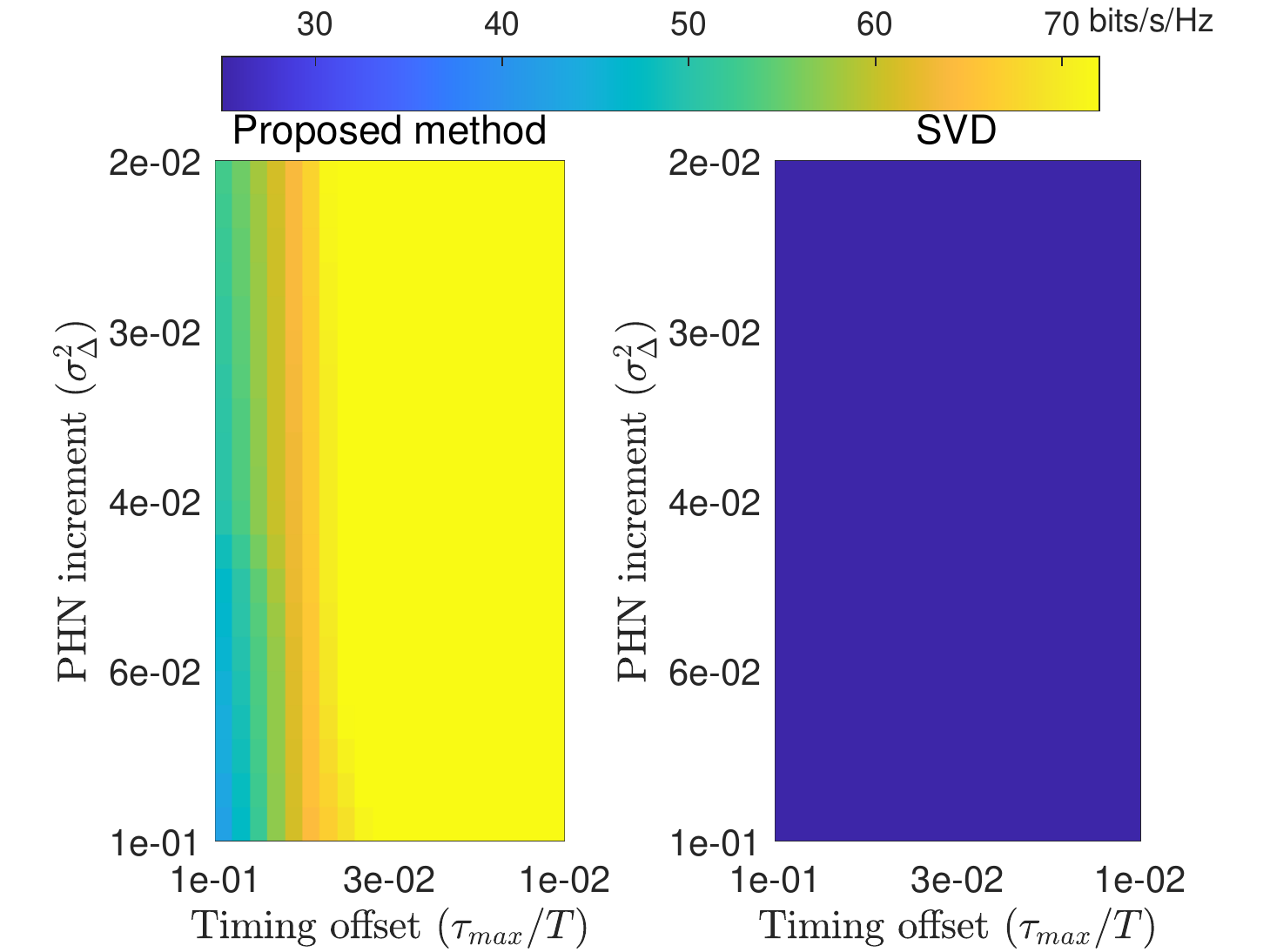}\caption{Theoretical sum-rate performance (Eq.(\ref{eq:Precprob})) of the
		proposed method and SVD under different physical impairments, $\rho=10$,
		SNR=$47$dB, $N=M=8$. A $30\times30$ grid is constructed for each
		algorithm with the performance averaged by 100 independent trials
		at each grid point. \label{fig:precperform}}
\end{figure}

To verify the robustness of the precoder/decorrelator, multiple simulations
are performed under different physical impairments. Specifically,
the received signal is generated by (\ref{eq:Symbol-Sig-Model}) considering
one sample per symbol with $\tau_{max}/T\in(0.001,0.1)$ and $\sigma_{\Delta}\in(0.001,0.1)$.
The estimated CSI $\bar{\mathbf{H}}$ required for the precoder/decorrelator
is obtained by the channel estimation in subsection \ref{subsec:Channel-Estimation}.
In the optimization problem for the proposed method, $\varpi$ is
set as the capacity under the $4096$-QAM and (SER) $<0.001$. Fig.
\ref{fig:precperform} presents the theoretical sum-rate performance
(Eq.(\ref{eq:Precprob})) comparisons between the proposed precoder/decorrelator
and the SVD solution. The results show that the proposed method outperforms
SVD in all the considered scenarios. 

\begin{figure}
	\begin{centering}
		\includegraphics[width=0.4\textwidth]{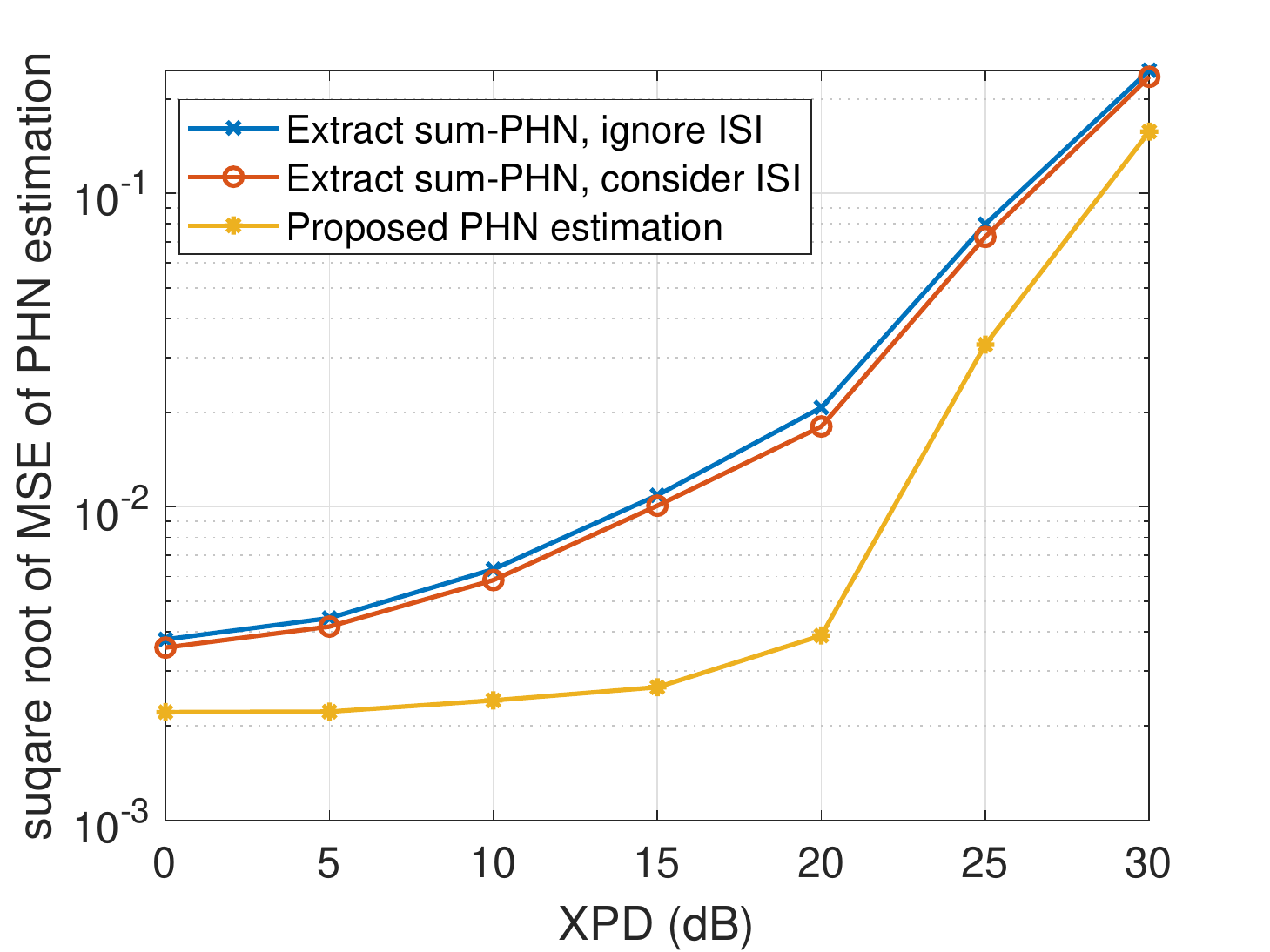}
		\par\end{centering}
	\caption{PHN estimation error in pilots over a frame.\label{fig:PHN-estimation-error}}
\end{figure}

Fig. \ref{fig:PHN-estimation-error} shows the performance of PHN
estimation in pilots over a frame for 100 Monte-Carlo trials. The
performance metric is the rooted MSE of the effective sum-PHN in each
MIMO link. The accumulated sum-PHNs at $q$-th pilot, denoted as vector
$\boldsymbol{\vartheta}\left[q\right]$, can also be
obtained from the per-antenna PHN $\boldsymbol{\phi}\left[q\right]$
via the transformation $\boldsymbol{\vartheta}\left[q\right]=\text{\ensuremath{\bar{\mathbf{I}}}}_{NM}\boldsymbol{\phi}\left[q\right]$.
The baselines extract the PHN by first getting an estimate of the
principal channel during the $q$-th pilot transmission $\hat{\mathbf{H}}_{q}\left[0\right]$,
then $\boldsymbol{\vartheta}\left[q\right]$ is estimated
by \cite{song2018design} 
\begin{equation}
\hat{\boldsymbol{\vartheta}}\left[q\right]=\angle\mathrm{vec}\left(\hat{\mathbf{H}}_{q}\left[0\right]\right)-\angle\mathrm{vec}\left(\hat{\mathbf{H}}\left[0\right]\right).\label{eq:Sum_PHN_Extract}
\end{equation}
In getting the principal channel, the first baseline ignores the ISI
and uses an LS algorithm \cite{mehrpouyan2012joint}\cite{song2018design}
$\hat{\mathbf{H}}_{q}\left[0\right]=\mathbf{Y}\left[q\right]\mathbf{X}^{H}\left(\mathbf{X}\mathbf{X}^{H}\right)^{-1}$
to directly estimate the principal channel, while the second baseline
uses the pilots to estimate the principal and ISI channels with the
LS-based channel estimation method introduced in Section \ref{subsec:Channel-Estimation},
and then extracts the desired principal channel $\hat{\mathbf{H}}_{q}\left[0\right]$.
As observed from Fig. \ref{fig:PHN-estimation-error}, for 0 dB <
XPD < 20 dB, the $\emph{proposed PHN estimation}$ method reaches
an estimation error at about 0.002 rad, while the errors of the two
$baselines$ increase rapidly to 0.01 rad as XPD increases. This is
because, with higher XPD, the phase of cross-polar links is more sensitive
to the MAI and noises due to small channel power in these links. It
is noted that at low XPD (around 0dB), the proposed PHN estimation
method still outperforms the baselines because it leverages the ISI
suppression capability of the precoder and decorrelator. The performance
of the proposed PHN estimation method starts to decay for high XPD
($>20$ dB) due to loss of measurements in the cross-polar links,
but it still outperforms the baselines.

\subsection{End-to-End Performance}

For the end-to-end performance evaluation, we perform 200 Monte-Carlo
trials (i.e., independent frames) with XPD $=20$ dB and $\rho=10$
dB in which our proposed holistic method is compared to several baselines.
These baselines partially adopt different techniques in the system:
\begin{itemize}
	\item \textbf{Baseline 1} \cite{song2018design}: This baseline first estimates
	the accumulated sum-PHN $\hat{\boldsymbol{\vartheta}}\left[q\right]$
	by (\ref{eq:Sum_PHN_Extract}). Then, the sum-PHN is converted back
	to the original transmitter and receiver PHN $\hat{\boldsymbol{\phi}}\left[q\right]$
	using the pseudo-inverse of $\text{\ensuremath{\mathbf{A}}}_{NM}$.
	Based on the information of $\hat{\boldsymbol{\phi}}\left[q\right]$,
	PHN de-rotation operations at both the receiver and the transmitter
	are applied before and after an equalizer to compensate per antenna
	PHN. The equalizer is calculated based on the principal channel estimated
	from the received preamble to suppress the MAI. Finally, an MMSE-FIR
	filter per stream is performed to suppress the ISI.
	\item \textbf{Baseline 2} \cite{mehrpouyan2012joint}: This baseline extracts
	accumulated sum-PHN $\boldsymbol{\vartheta}\left[q\right]$
	in the estimated channel matrix during pilot transmission by (\ref{eq:Sum_PHN_Extract}).
	This sum-PHN is then combined with the principal channel matrix estimated
	during the preamble transmission. An MMSE linear decorrelator is constructed
	based on the combined channel to equalize the effect of the PHN and
	channel gain. However, since this method pays no attention to the
	coherence loss effect and the ISI, thus it shows an inferior performance.
	\item \textbf{Baseline 3} (SVD): This baseline uses the unitary matrices
	of the SVD on the estimated principal channel as the precoder and
	decorrelator. Because this method ignores the ISI and MAI, thus it
	suffers a poor performance in our considered scenario.
\end{itemize}

In this experiment, channel coding is not considered and all the approaches
adopt the same timing synchronization. To perform an end-to-end experiment
and show reasonable results, we take an adaptive modulation strategy
by estimating the SINR via Eq. \ref{eq:SINR} before data transmission.
After the precoder/decorrelator calculation, modulation levels according
to the estimated SINR are suggested according to the relationship
between the modulation level, SINR, and SER where we fix SER $<0.001$
for each stream. Then, the data is transmitted with the Gray mapped
$Q_{M}$-QAM symbol where $Q_{M}$ is the suggested modulation level
and is demodulated with the hard decision. Note that practical adaptive
modulation may not be operated in this way, but taking this strategy
will not influence the performance comparisons. 

\begin{figure}
	\begin{centering}
		\includegraphics[scale=0.5]{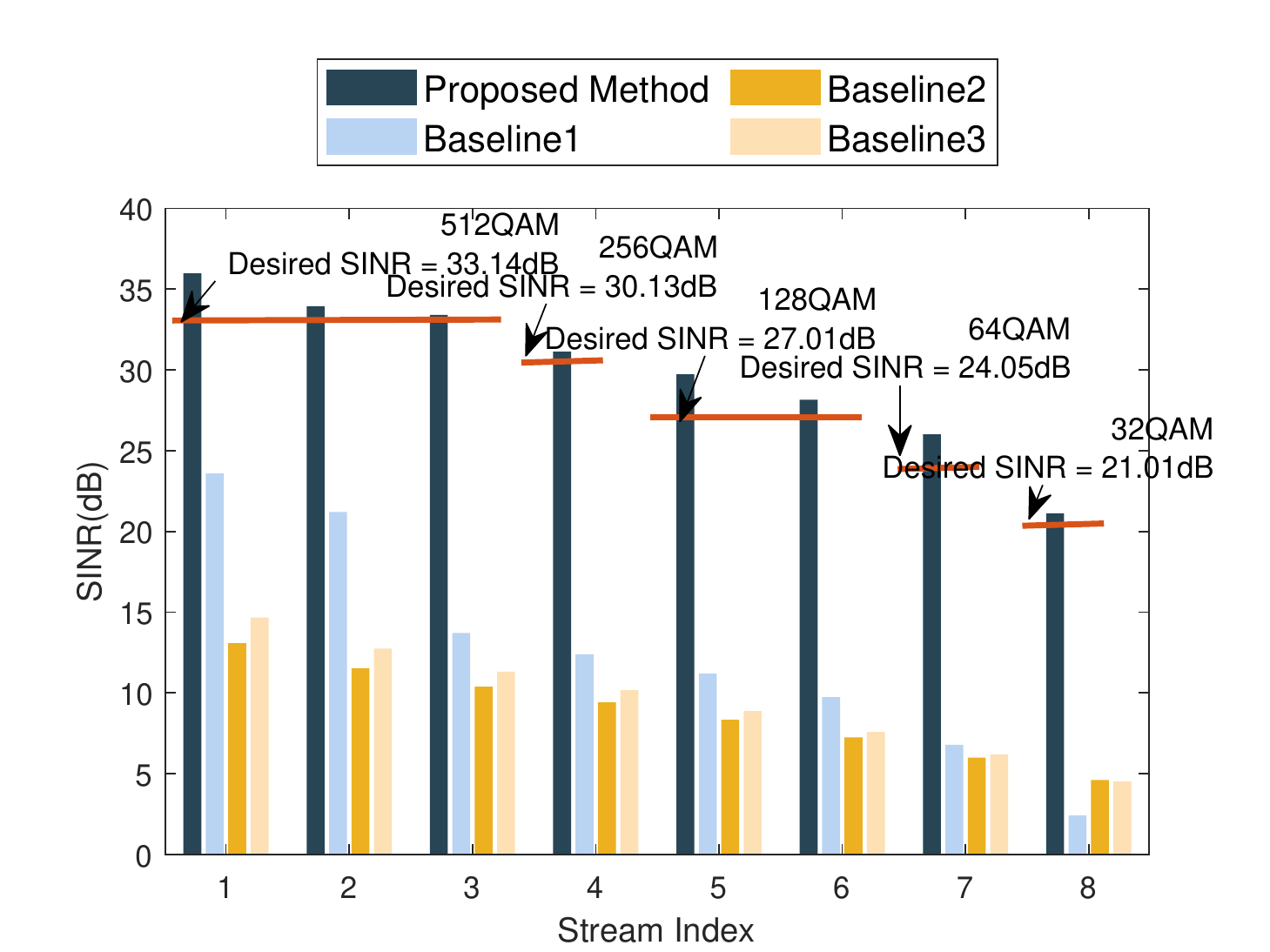}
		\par\end{centering}
	\caption{Per-stream SINR performance in the End-to-End experiment.\label{fig:E2Esinr}}
\end{figure}

To verify the superiority of the proposed scheme, we first show per
stream SINR to verify the reliability and efficiency of the system
in a detailed view. Specifically, for the $m$-th stream, $\text{SINR}_{m}(dB)=10\log_{10}\mathbb{E}_{k}\left[\frac{||\mathbf{w}_{m}^{H}\left(e^{j\mathrm{diag}\left(\Delta\boldsymbol{\phi}^{rx}\left[k\right]\right)}\mathbf{H}\left[0\right]e^{j\mathrm{diag}\left(\Delta\boldsymbol{\phi}^{tx}\left[k\right]\right)}\right)\mathbf{s}\left(k\right)||_{2}^{2}}{||r_{m}(b)-\mathbf{w}_{m}^{H}\left(e^{j\mathrm{diag}\left(\Delta\boldsymbol{\phi}^{rx}\left[k\right]\right)}\mathbf{H}\left[0\right]e^{j\mathrm{diag}\left(\Delta\boldsymbol{\phi}^{tx}\left[k\right]\right)}\right)\mathbf{s}\left(k\right)||_{2}^{2}}\right]$.
In Fig. \ref{fig:E2Esinr}, per stream SINR performance is given among
different methods. The highest QAM modulation level that can be achieved
by the proposed method at each stream is also given. It is obvious
that the proposed method outperforms the baselines of each stream
with an over 10 dB gain, and the modulation level achieved by the
proposed method can not be afforded by baselines.
To show an overall performance, we then compare the \emph{BER} and
\emph{Spectrum Efficiency} in Table. \ref{tab:End-to-End-BER-and}.
Note that both results are averaged over uplink and downlink effective
transmissions.
\begin{table}
	\caption{End-to-End BER and Spectrum Efficiency performance.\label{tab:End-to-End-BER-and}}
	
	\centering{}%
	\begin{tabular}{ccc}
		\toprule 
		& \multirow{2}{*}{\textbf{BER($10^{-3}$)}}&\textbf{Spectrum Efficiency}\\
		&~&\textbf{(bits/s/Hz)}\tabularnewline
		\midrule
		\midrule 
		Proposed & 0.2221 & 60\tabularnewline
		\midrule 
		Baseline 1 & 0.6740 & 15\tabularnewline
		\midrule 
		Baseline 2 & 0.2036 & 8\tabularnewline
		\midrule 
		Baseline 3 & 0.4401 & 8\tabularnewline
		\bottomrule
	\end{tabular}
\end{table}
From Table. \ref{tab:End-to-End-BER-and}, the results show that the
proposed method can achieve an over $4$ times spectrum efficiency
gain compare to all the baselines with a nearly identical BER.

\section{Conclusion\label{sec:Conclusion}}

This paper proposes a holistic solution to achieve high spectral efficiency
for an LoS MIMO system equipped with dual-polarized antenna arrays
for wireless backhaul using FDD. Various physical impairments such
as TOs and phase noises as well as the effect of multipath are taken
into account. An LS-based TO estimator is proposed with a new set
of preamble sequences to perform timing synchronization. After channel
estimation, an optimization-driven precoder/decorrelator design is
proposed to suppress the MAI and ISI introduced by the multipath to
maximize the theoretical sum-rate. Additionally, a PHN estimation,
tracking, and compensation scheme is proposed to tackle the problem
of the coherence loss of precoder/decorrelator due to the drifting
of the PHN over a frame. With a simulation setup in a $8\times8$
LoS MIMO demonstrator, the proposed method can achieve high spectral
efficiency at 60 bits/s/Hz. Simulation results also show that our
proposed solution outperforms the existing baselines.

\begin{appendices}
	
	\section{\label{subsec:PRE} Derivation of problem (\ref{eq:Precprobapp})}
	
	The derivation is similar to \cite[Appendix A]{shi2011iteratively},
	except for the procedure used to approximate the non-smoothness in
	the original problem (\ref{eq:Precprob}). First, when fixing the
	other variables, the optimal decorrelator, $\ensuremath{\mathbf{\tilde{W}}}$
	is the MMSE decorrelator as shown in (\ref{eq:upW-1}). Second, to
	update the auxiliary variable $\boldsymbol{\Gamma}\in\mathbb{C}^{N_{s}\times N_{s}}$,
	we first check the first-order optimality of $\boldsymbol{\Gamma}$
	and then do the projection by considering the box constraint for it.
	The optimal $\boldsymbol{\Gamma}$ is a diagonal matrix with $\boldsymbol{\Gamma}_{m,m}=\min\{[\mathbf{E}(\ensuremath{\mathbf{\tilde{W}}},\mathbf{F})]_{m,m}^{-1},2^{\text{\ensuremath{\varpi}}}\}$.
	Substituting the optimal $\ensuremath{\mathbf{\tilde{W}}}$ and $\boldsymbol{\Gamma}$
	into (\ref{eq:Precprob}) we have 
	\begin{align*}
	\underset{\mathbf{F}}{{\mathrm{maximize}}}\quad & -\sum\limits _{m=1}^{N_{s}}\min\{[\mathbf{E}(\ensuremath{\mathbf{\tilde{W}}(\mathbf{F})},\mathbf{F})]_{m,m}^{-1},2^{\text{\ensuremath{\varpi}}}\}[\mathbf{E}(\mathbf{\tilde{W}}(\mathbf{F}),\mathbf{F})]_{m,m}\\
	& +\sum\limits _{m=1}^{N_{s}}\min\{\log(1+\text{SINR}_{m}),2^{\text{\ensuremath{\varpi}}}\}\\
	& s.t\quad Tr(\mathbf{F^{H}F})\leq P,
	\end{align*}
	where $-\sum\limits _{m=1}^{N_{s}}\min\{[\mathbf{E}(\ensuremath{\mathbf{\tilde{W}}(\mathbf{F})},\mathbf{F})]_{m,m}^{-1},2^{\text{\ensuremath{\varpi}}}\}[\mathbf{E}(\mathbf{\tilde{W}}(\mathbf{F}),\mathbf{F})]_{m,m}$
	is a constant when $\min\{[\mathbf{E}(\ensuremath{\mathbf{\tilde{W}}(\mathbf{F})},\mathbf{F})]_{m,m}^{-1},2^{\text{\ensuremath{\varpi}}}\}=[\mathbf{E}(\ensuremath{\mathbf{\tilde{W}}(\mathbf{F})},\mathbf{F})]_{m,m}^{-1},\forall m$.
	When a stream has a very small MMSE (or high SINR), there is truncation,
	i.e., $\min\{[\mathbf{E}(\ensuremath{\mathbf{\tilde{W}}(\mathbf{F})},\mathbf{F})]_{m,m}^{-1},2^{\text{\ensuremath{\varpi}}}\}=2^{\text{\ensuremath{\varpi}}}$.
	However, this truncation usually causes a small error when $2^{\text{\ensuremath{\varpi}}}$is
	very large, since in this case, $[\mathbf{E}(\mathbf{\tilde{W}}(\mathbf{F}),\mathbf{F})]_{m,m}$
	is very small.
	
	\section{Derivation of problem (\ref{eq:PHN_LS})\label{subsec:DerivationPHNLS}}
	
	Elementwisely, the $\left\{ m,k\right\} $-th element of $\mathbf{R}\left[q+1\right]$
	can be expressed as{\small{}
		\begin{align}
		& r_{m}\left(k|q+1\right)\nonumber \\
		= & \sum_{d=-D}^{D}\sum_{i=1}^{N}\sum_{j=1}^{M}W_{im}^{*}\left(d\right)h_{ij}\left[-d\right]x_{j}\left(k\right)e^{j\left(\Delta\phi_{i}^{rx}\left[q\right]+\Delta\phi_{j}^{tx}\left[q\right]\right)}+v_{m}\left(k\right)\nonumber \\
		\approx & \sum_{d=-D}^{D}\sum_{i=1}^{N}\sum_{j=1}^{M}\alpha_{m,i,j}^{d}\left(k\right)\left[1+j\left(\Delta\phi_{i}^{rx}\left[q\right]+\Delta\phi_{j}^{tx}\left[q\right]\right)\right]+v_{m}\left(k\right),\label{eq:Per-Stream-SigMod}
		\end{align}
	}where $\Delta\phi_{i}^{rx}\left[q\right]$ and $\Delta\phi_{j}^{tx}\left[q\right]$
	represents the $i$-th and $j$-th element of $\Delta\boldsymbol{\phi}^{rx}\left[q\right]$
	and $\Delta\boldsymbol{\phi}^{tx}\left[q\right]$, respectively,
	$x_{j}\left(k\right)=\left[\mathbf{X}\right]_{j,k}$ is the $k$-th
	transmitted signal at the $j$-th transmit antenna, and we define
	$\alpha_{m,i,j}^{d}\left(k\right)\triangleq W_{im}^{*}\left(d\right)h_{ij}\left[-d\right]x_{j}\left(k\right)$.
	In (\ref{eq:Per-Stream-SigMod}), we used first-order Taylor expansion
	$e^{j\Delta\phi}\approx1+j\Delta\phi$ for small PHN variation $\Delta\phi$
	for the approximation.
	
	As seen from (\ref{eq:Per-Stream-SigMod}), each element $r_{m}\left(k|q+1\right)$
	is an observation for the $M+N$ PHN parameters, but each transmitted
	signal is distorted by the sum-PHN process $\Delta\phi_{i}^{rx}\left[q\right]+\Delta\phi_{j}^{tx}\left[q\right]$.
	The $M\times N$ sum-PHN can be calculated in principle if we have
	adequate observations, but the transformation back to their original
	form $\Delta\boldsymbol{\phi}\left[q\right]=\left[\left(\Delta\boldsymbol{\phi}^{rx}\left[q\right]\right)^{T},\left(\Delta\boldsymbol{\phi}^{tx}\left[q\right]\right)^{T}\right]$
	is impossible \cite{hadaschik2005improving} as we lack a reference
	phase. To eliminate the degree of freedom, we can just fix $\Delta\phi_{1}^{rx}\left[q\right]=0$
	while preserving the effective sum-PHN in each link.
	
	Using matrix representation, we define the matrix $\left[\boldsymbol{\Omega}_{m}^{d}\left(k\right)\right]_{i,j}=\alpha_{m,i,j}^{d}\left(k\right)$,
	the vector $\boldsymbol{\xi}_{m}\left(k\right)$ and $\boldsymbol{\eta}_{m}\left(k\right)$
	in (\ref{eq:vector-define}) and (\ref{eq:eta-define}), respectively,
	and the approximation in (\ref{eq:Per-Stream-SigMod}) can be rewritten
	as
	\begin{align*}
	& r_{m}\left(k|q+1\right)-\mathbf{1}_{MN\left(2D+1\right)}^{T}\boldsymbol{\eta}_{m}\left(k\right)\\
	\approx & \boldsymbol{\xi}_{m}^{T}\left(k\right)\Delta\boldsymbol{\phi}\left[q\right]+v_{m}\left(k\right),\ \forall m=1,...,N_{s},k=1,...,L_{p},
	\end{align*}
	where $\bar{\mathbf{I}}_{NM}$ is defined in Eq.\ref{eq:Adef}. Stacking
	the observations into a big column vector during the $\left(q+1\right)$-th
	pilot transmission, we shall arrive at the LS problem as (\ref{eq:PHN_LS}). 
\end{appendices}

\section*{About the Authors}\footnotesize\vskip 2mm

\bibliographystyle{unsrt}
%% argument is your BibTeX string definitions and bibliography database(s)
\bibliography{reference}

\end{document}